\documentclass[conference,compsoc]{IEEEtran}
\IEEEoverridecommandlockouts

\usepackage{paralist}
\usepackage{grffile}

\usepackage[absolute,showboxes]{textpos}

\usepackage{balance}

\newcommand\new[1]{#1} 

\newcommand{\ie}{\textit{i}.\textit{e}.~}
\newcommand{\eg}{\textit{e}.\textit{g}.,~}

\newcommand{\eat}[1]{}

\usepackage{tabularx}
\newcolumntype{L}[1]{>{\raggedright\arraybackslash}p{#1}}
\newcolumntype{C}[1]{>{\centering\arraybackslash}p{#1}}
\newcolumntype{R}[1]{>{\raggedleft\arraybackslash}p{#1}}

\usepackage{graphicx}
\usepackage{blindtext}
\usepackage[hidelinks]{hyperref}
\usepackage{multicol,multirow}
\usepackage{multirow}
\usepackage{booktabs}
\usepackage{colortbl}
\usepackage{color}
\usepackage[usernames,dvipsnames,svgnames,table]{xcolor}
\usepackage{xargs}                      % Use more than one optional parameter in a new commands
\usepackage[compact]{titlesec}

\usepackage[colorinlistoftodos,prependcaption,textsize=tiny]{todonotes}

\usepackage{pifont}

\newcommand{\cmark}{\color{ForestGreen} {\ding{51}}}%
\newcommand{\xmark}{\color{BrickRed}{\ding{55}}}%

\newcommandx{\am}[2][1=]{\todo[linecolor=red,backgroundcolor=red!25,bordercolor=red,#1]{#2}}
\begin{document}

\title{Protected or Porous:\\A Comparative Analysis of Threat Detection Capability of IoT Safeguards}

%\author{\IEEEauthorblockN{Anonymous Authors}}
\author{\IEEEauthorblockN{Anna Maria Mandalari}
\IEEEauthorblockA{\textit{University College London} \\
\textit{a.mandalari@ucl.ac.uk}}
\and
\IEEEauthorblockN{Hamed Haddadi}
\IEEEauthorblockA{\textit{Imperial College London} \\
\textit{h.haddadi@imperial.ac.uk}
\and
\IEEEauthorblockN{Daniel J. Dubois, David Choffnes}
\IEEEauthorblockA{\textit{Northeastern University} \\
\textit{\{d.dubois,d.choffnes\}@northeastern.edu}}}}
\maketitle

\setlength{\TPHorizModule}{\paperwidth}
\setlength{\TPVertModule}{\paperheight}
\TPMargin{5pt}
\begin{textblock}{0.8}(0.1,0.02)
  \noindent
  \footnotesize
  If you cite this paper, please use the IEEE reference:
  Anna Maria Mandalari, Hamed Haddadi, Daniel J. Dubois, David Choffnes.
  Protected or Porous: A Comparative Analysis of Threat Detection Capability of IoT Safeguards.
  In \textit{2023 IEEE Symposium on Security and Privacy (Oakland).}
\end{textblock}

\begin{abstract}
%\boldmath
%!TEX root = ../paper.tex
Consumer Internet of Things (IoT) devices are increasingly common, from smart speakers to security cameras, in homes.
Along with their benefits come potential privacy and security threats. 
To limit these threats a number of commercial services have become available (IoT safeguards). 
The safeguards claim to provide protection against IoT privacy risks and security threats. 
However, the effectiveness and the associated privacy risks of these safeguards remains a key open question.
In this paper, we investigate the threat detection capabilities of IoT safeguards for the first time.
We develop and release an approach for automated safeguards experimentation to reveal their response to common security threats and privacy risks.
We perform thousands of automated experiments using popular commercial IoT safeguards when deployed in a large IoT testbed.
Our results indicate not only that these devices may be ineffective in preventing risks, but also their cloud interactions and data collection operations may introduce privacy risks for the households that adopt them.
\end{abstract}

%!TEX root = ../paper.tex
\section{Introduction}
\label{sec:introduction}

Consumer Internet of Things (IoT) devices, including smart home assistants, smart TVs, surveillance cameras, and connected health monitoring appliances, are increasingly seen in and around homes.
While these devices come with interesting and beneficial services, they also expose their users to privacy and security risks, such as
personal information exposure, misactivations, or being hit by malware and involuntarily becoming part of botnets~\cite{moniotr, 10.1145/3319535.3354198, dubois2020speakers}.

Recently, a number of commercial \emph{antivirus}-like hardware routers and software services (safeguards) have become available (see Section~\ref{sec:safeguards}), claiming to protect against IoT threats by blocking malicious or otherwise undesirable connections. 
These devices and services often promise to provide detection and threat mitigation against network intrusions, malware and anomalies, and/or exposure of Personally Identifiable Information (PII). %bad sentence by me (hamed), needs a rewrite 
%However, these solutions often rely on users to configure the safeguard each device, and often provide an \textit{all-or-nothing} connectivity option for traffic destinations without considering whether blocking traffic will break device functionality.  
However, the blackbox nature of the safeguards, the complicated nature of IoT devices, their associated apps and services, and lack of availability of IoT virus/malware databases, makes it challenging to independently and rigorously measure the effectiveness of these safeguards against their claims.

In this paper, we perform the first exhaustive and large-scale investigation of the functionality and effectiveness of commercial IoT safeguards and their services. 
We use our large-scale  IoT testbed, along with several smartphones, to launch over 4,000 automated attacks, anomalies, security and privacy threats against 79 IoT devices protected by the safeguarding systems. 
%We perform thousands of automated attacks over two months (using a variety of attack mechanisms and  approaches) against popular consumer IoT devices in our testbed, protected by safeguards. 
The security attacks, privacy threats, and vulnerabilities that we evaluate include anomalies in behavior and traffic, network attacks (\eg port scanning, SYN/UDP/DNS flooding, \emph{etc}.), and PII exposure and unencrypted traffic assessment. 

Surprisingly, we find that these commercial safeguards today provide very little protection and coverage across widely-available, well-known, and documented attacks and privacy risks. 
Our assessment of the safeguard services, and their cloud providers and third party interactions, also reveals the privacy risks associated with using these services and the PII  shared by them with third parties.
We demonstrate various technical shortcomings of these services in fulfilling their promised protections, and provide insights into the overheads and side effects of using these safeguards.

Our key research contributions include:
\begin{compactitem}
\item We develop an automated methodology for benchmarking existing commercial IoT security and privacy safeguards using large-scale, diverse experiments and set of anomalies and attacks;
\item We assess the (in)effectiveness of popular safeguards against existing network and device anomaly attacks, and in identifying privacy risks;
\item We evaluate their overhead in terms of traffic;
\item and finally, we study the privacy policies associated with these safeguards to assess the privacy risks and implications of using these (mostly) cloud-based services and their third-party dependencies.
\end{compactitem}

In summary, we find that existing commercial systems for IoT protection fall exceedingly short of vendor claims, and likely user expectations. We argue for increased transparency and accountability in this space, given the cost of the services and risks when unprotected IoT devices are compromised.  
%Our code and data for the experiments will be made publicly available with the public release of this paper.
\new{We make our experiment software and datasets available at \url{https://iotrim.github.io/safeguards.html}.}

\noindent \textbf{Responsible Disclosure}. We responsibly disclosed our results with 
the safeguard manufacturers in this study. We received responses from five manufacturers; 
%who indicated 
%that they would investigate their shortcomings in detecting the threats in our study. 
\new{three} of them provided clarifications. 
We include these responses (with permission) in Section~\ref{sec:discussion} and in the Appendix. 
%We will include details of any subsequent feedback and mitigations in the final version of this paper. We are currently working with a Data Protection Authority to responsibly disclose the issues found.

%!TEX root = ../paper.tex
\section{\new{Assumptions, Goals, and Challenges}}
\label{sec:assumptions}

%In this section we summarize the threat model and goals of this work.

%\subsection{Definitions}
%\label{sub:definitions}
%
%\noindent \textbf{Safeguards.} We define safeguards as solutions to secure an IoT connected home, offering an interface for protecting the consumer IoT devices. 
%These services may receive all data from the IoT devices when they are activated.
%We consider 8 commercial safeguards listed in Section~\ref{sec:testbed}
%
%\noindent \textbf{Notifications.} We define notification every time a safeguard notifies the user for a potential privacy or security threat. This may or may not be followed by sending data over the Internet to its cloud-enabled safeguard. 

\textbf{Threat Model.}
%\subsection{Threat Model}
%\label{sub:threat}
We consider the following threat model. 

\noindent \emph{Victim.} The victim is any person that owns, uses, or benefits from consumer IoT devices under safeguard protection.

\noindent \emph{Adversary.} The adversary is any party that can access the IoT device traffic. Examples include external privacy and security threats, malicious IoT devices, safeguards.

\noindent \emph{Threat.} We assume a router is deployed in a smart home. We are broadly concerned with the
ability of ISPs, IoT manufactures, on-path network observers, VPN
service providers, and third-parties to infer user in-home activities
from smart home network traffic rate traces. The network traffic
rate metadata, including inbound/outbound traffic rates, network protocols, source, and destination IPs, package sizes etc., are
accessible to many entities. Adversaries may be
incentivized to infer user activities where users do
not want to share their privacy-sensitive information with them. 
In particular, we consider two types of threats: Security (Mirai, Scan, \emph{etc}.~\cite{antonakakis2017understanding,kolias2017ddos}), Privacy (IoT devices can potentially expose information about their users~\cite{moniotr}).

%\begin{itemize}
%\item \emph{Security.} Most IoT devices are vulnerable to security attacks (Mirai, Scan, etc.~\cite{antonakakis2017understanding,kolias2017ddos}).
%\item \emph{Privacy.} IoT devices can potentially expose information about their users~\cite{moniotr}.
%
%\end{itemize}

\noindent \textbf{Goals.}
%\subsection{Goals \new{and Challenges}}
\label{sub:goals}
The main goal of this work is to analyze the effectiveness of consumer IoT safeguards. In particular, this work answers the following research questions (RQ):

\noindent \emph{RQ1. What are the privacy and security implications on how a safeguard works?} 
Our goal is to characterize how safeguards operate (i.e., ARP spoofing, packet inspection, destinations contacted) to see if they produce security or privacy concerns. In particular,  we consider if they need the cloud to operate or if they operate locally. We also check if they use third-party services to operate. To address this, we propose a testbed for systematically studying the safeguards and tracking their network traffic (Section~\ref{sec:rq1}).

\noindent \emph{RQ2. Do the safeguards detect threats?}  
Safeguards notify the user when detecting privacy or security threats. We check their capability to do so (Section~\ref{sec:rq2}). 

\noindent \emph{RQ3. What are the side effects of the safeguards?} 
We measure whether the safeguards add traffic overhead. We also check whether they overprotect the user (i.e., we analyze false positive threat detection), and the implications of privacy policy disclosures (Section~\ref{sec:rq3}). 

%\subsection{Non-Goals}
%\subsection{Non-Goals}
%\label{subsec:nongoals}
\noindent \textbf{Non-Goals.}
In this work, we do not consider the following as goals, and leave them for future work.

\noindent \emph{No control over how a safeguard works internally.}
We consider the safeguard as a blackbox that provides a finite set of threat detection capabilities and
that communicates over the Internet. For these safeguards,
we have no control over their internal functionality, but we
can still interact with them using their user interface (i.e., companion app)
and we can measure their network activity.
%
%While we consider the visibility of the content out of scope, we
%are able to see the destinations of such traffic. We make
%this assumption because 
The vast majority of the safeguards' traffic is encrypted and we assume that we cannot install custom
self-signed certificates to use man-in-the-middle techniques to intercept their encrypted traffic.

\noindent \emph{We do not test all threats.}
Safeguards typically offer several threat detection capabilities; however, we apply
our methodology only on a subset of them for
every safeguard under test so that we can cover the same threats for different safeguards. To be included in our list, a threat must be declared as detectable by at least three safeguards.

\noindent \emph{Consumer IoT devices.}
We focus on safeguards and devices that target consumers; hence, the enterprise environment is out of scope for this paper.

%In the next section, we present details of the methods we use for gathering data to answer these questions (Section~\ref{sec:testbed},~\ref{sec:method}). We then analyze the data to answer our research questions in Section~\ref{sec:rq1_evaluation},~\ref{sec:rq2_evaluation},~\ref{sec:rq3_evaluation}, and provide a discussion of the results in Section~\ref{sec:discussion}.

\noindent \textbf{\new{Challenges.}}
\new{The main challenge we overcome in this paper is the creation of a methodology that enables us to automate the testing of commercial IoT safeguards. The IoT safeguards operate in closed systems, making it difficult to test them in a controlled environment. To overcome this, we create a large IoT testbed used to simulate real-world scenarios. Our methodology is designed to address the challenges in analyzing blackbox solutions, where we have limited visibility on how the system works internally and the manufacturers provide no open, repeatable methodology for the evaluation of their claims.}

%!TEX root = ../paper.tex
\section{Safeguards Overview}
\label{sec:safeguards}

%In this section we summarize the characteristics of the safeguards we test.

%\subsection{Safeguards}
%\label{sec:safeguards_name}

 \begin{figure}[t!]
  \centering
  \includegraphics[width=0.75\columnwidth]{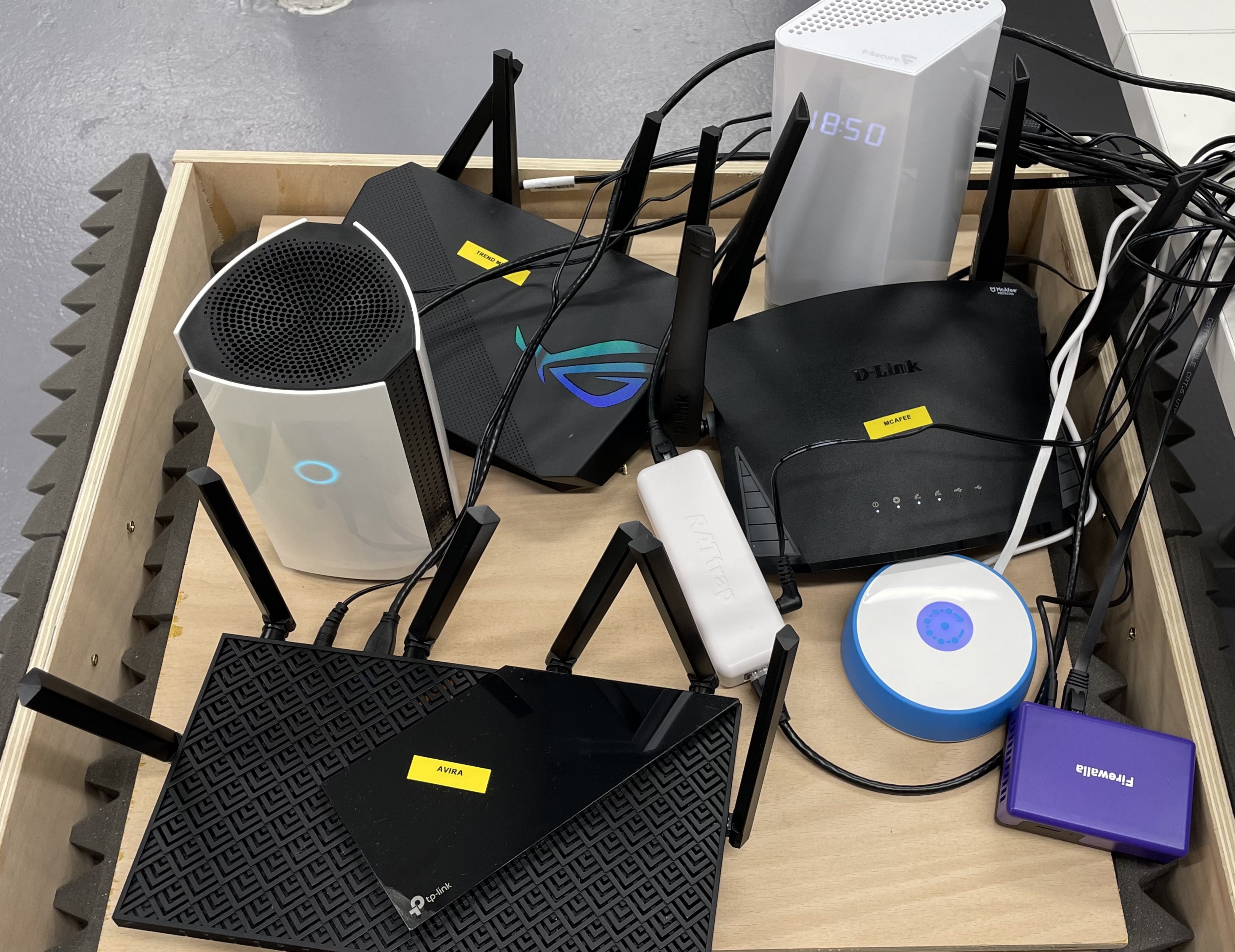}
  \vspace{-0.15cm}
  \caption{Picture of the safeguards testbed in our lab.}
  \label{fig:setup_lab}
\end{figure}

\textbf{Safeguards tested.}
We consider the 8 most popular safeguards in the European market (see Figure~\ref{fig:setup_lab}). To assess popularity, we use sources commonly available to a user, such as Amazon and Google rankings. The safeguards we consider are the following:

\noindent \emph{Avira.} Avira offers its services through the TP-Link commercial router. The service is called \textit{HomeShield}~\cite{avira} and it is meant to secure the users' connected devices against cyberattacks and various online threats.

\noindent \emph{Bitdefender.} Bitdefender offers its services through the \textit{Bitdefender BOX}~\cite{bitdefender}, a cybersecurity ecosystem for the home network and family devices; this device acts as a router.  %Bitdefender is also available as a software add-on for certain commercial NETGEAR routers. %commercial router. The box acts as a router. 

\noindent \emph{F-Secure.} F-Secure offers its services through the \textit{SENSE security router}~\cite{fsecure}, blocking malicious websites and other threats, and securing smart devices against cyber attacks.

\noindent \emph{Fingbox.} Fingbox~\cite{fing} connects directly to the home router. It keeps users' Internet safe and secure by automatically blocking intruders, hackers and unknown devices. 

\noindent \emph{Firewalla.} Firewalla is a smart firewall for privacy protection~\cite{firewalla} and cyber security. The code is open source. For our analysis we select \textit{Firewalla purple}. %the one with the most advanced features. 

\noindent \emph{McAfee.} McAfee offers its services through the D-Link commercial router. The service is called \textit{Secure Home Platform}~\cite{mcafee}. McAfee global threat intelligence is a cybersecurity service that identifies and blocks emerging threats. %\textit{Secure Home Platform} is offered for free.

\noindent \emph{RATtrap.} RATtrap offers its service through a box connected between the ISP and the home router~\cite{rattrap}. They offer universal protection and instant ad blocking. %The company behind RATtrap is \textit{IoT Defense}~\cite{iotdef}, a security software development company based in US.

\noindent \emph{TrendMicro.} TrendMicro offers its services through the Asus commercial router~\cite{trendmicro}. ASUS \textit{AiProtection} offers protection for home connected devices. % from compromise, and unhealthy Internet usage.

%\subsection{Safeguards Types}
%\label{sub:safeguardtypes}

 \begin{figure}[t!]
  \centering
  \includegraphics[width=0.8\columnwidth]{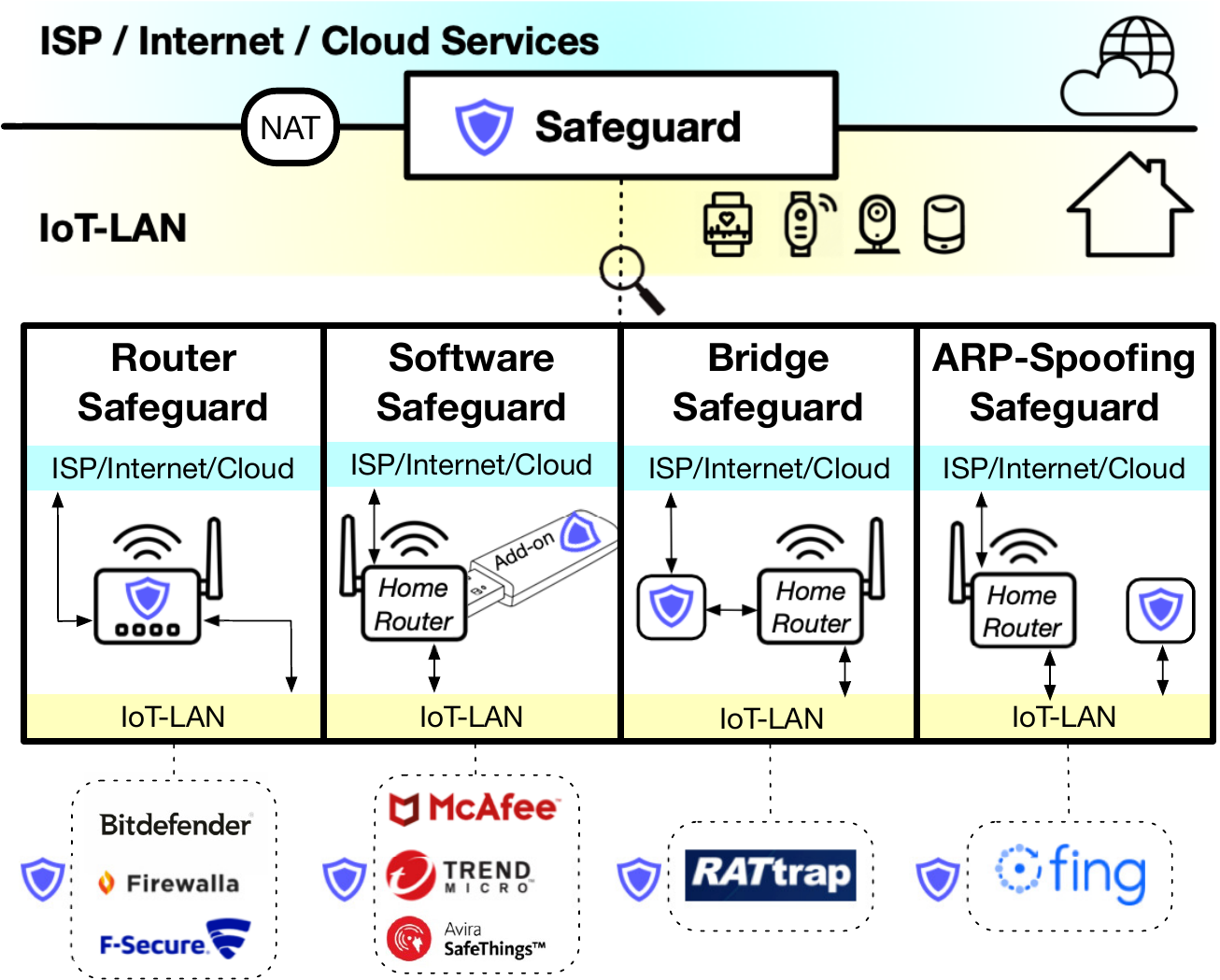}
    \vspace{-0.15cm}
  \caption{Safeguard setup overview and safeguard types.}
  \label{fig:setup_safeguard}
\end{figure}

\noindent \textbf{Safeguards types.}
We analyze four types of safeguards, as shown in Figure~\ref{fig:setup_safeguard}.

\noindent \emph{Router safeguard:} a physical safeguard that behaves as a router. It has two network interfaces, one connected to the Internet and the other connected to the local network hosting the IoT devices to protect.

\noindent \emph{Software safeguard:} a software add-on that can be installed on a compatible commercial router already owned by the user, to transform it into a router safeguard. %Once the user install the safeguard add-on, this type of safeguard is not distinguishable from a router safeguard.

\noindent \emph{Bridge safeguard:} a physical safeguard that behaves as a two-port bridge, one port connected to the Internet and the other designed to be connected to a user-provided router. The disadvantage of this setup when compared to the router safeguard is the need for an additional router and the fact that the safeguard has no visibility into the MAC addresses of the IoT devices connected to the router. %When used in tandem with a router, from the user perspective this safeguard is equivalent to a router safeguard.

\noindent \emph{ARP-Spoofing safeguard:} a physical firewall with only one network port that is designed to be connected to the IoT-LAN. Unlike the other types of safeguards, the ARP-Spoofing type is not a conventional router and it does not need to be interposed between the router and the Internet to analyze the traffic routed between the IoT-LAN and the Internet. Instead, it uses spoofed ARP packets to make the IoT devices believe it is the router, and then forwards such packets to the real router after intercepting them. 
%Since this safeguard still intercepts the traffic between the IoT-LAN and the Internet, we consider it functionally equivalent to a router safeguard, when paired to a router.

%!TEX root = ../paper.tex

\begin{table*}[t!]
\centering
	\caption{Summary of the characteristics of the safeguards}
	\label{tab:safeguards}
	\vspace{-2mm}
	\resizebox{0.9\textwidth}{!}{
		\centering
		\begin{tabular}{p{0.13\textwidth}cccccccc}
		\centering
			\textbf{Property} & \textbf{Avira} & \textbf{Bitdefender} & \textbf{F-Secure}& \textbf{Fingbox}& \textbf{Firewalla}& \textbf{McAfee}& \textbf{RATtrap}& \textbf{TrendMicro}\\ 
			\midrule
%General
\textbf{Type} & Software  &Router & Router &ARP-Spoofing & Router&Software &Bridge & Software   \\
\cline{1-9}
\textbf{Control} &  App &App & App & App& App& App& App& App   \\
\cline{1-9}
\textbf{Maintainer Type} &  Commercial &Commercial & Commercial & Commercial& Commercial& Commercial& Commercial& Commercial   \\
\cline{1-9}
\textbf{Open Source} &  \xmark &\xmark & \xmark & \xmark& \cmark& \xmark& \xmark& \xmark   \\
\cline{1-9}
\textbf{Security} &  \cmark &\cmark & \cmark & \cmark& \cmark& \cmark& \cmark& \cmark   \\
\cline{1-9}
\textbf{Privacy} &  \cmark &\cmark & \xmark & \xmark& \cmark& \xmark& \xmark& \xmark   \\
\cline{1-9}
\textbf{Business Model} &  Subscription &Subscription & Subscription & One-off & One-off& One-off& Subscription& One-off    \\
\cline{1-9}
\textbf{Price} &  \$54.99y &\$99y & \$99.99y & \$99& \$329.00& Free & \$108y& Free \\
\bottomrule	
		\end{tabular}
	}
\end{table*}

Table~\ref{tab:safeguards} summarizes the characteristics  of each safeguard. We report the type of safeguard (Type); the way the safeguard can be controlled (Control); how the safeguard is maintained (Maintainer Type); if the safeguard code is open source (Open Source); if the safeguard protects from security (Security) or privacy (Privacy) threats; if the service is offered by subscription or it is a one-off payment (Business Model); the price of the service (Price).

%!TEX root = ../paper.tex
\section{Testbed}
\label{sec:testbed}

 \begin{figure}[t!]
  \centering
  \includegraphics[width=0.9\columnwidth]{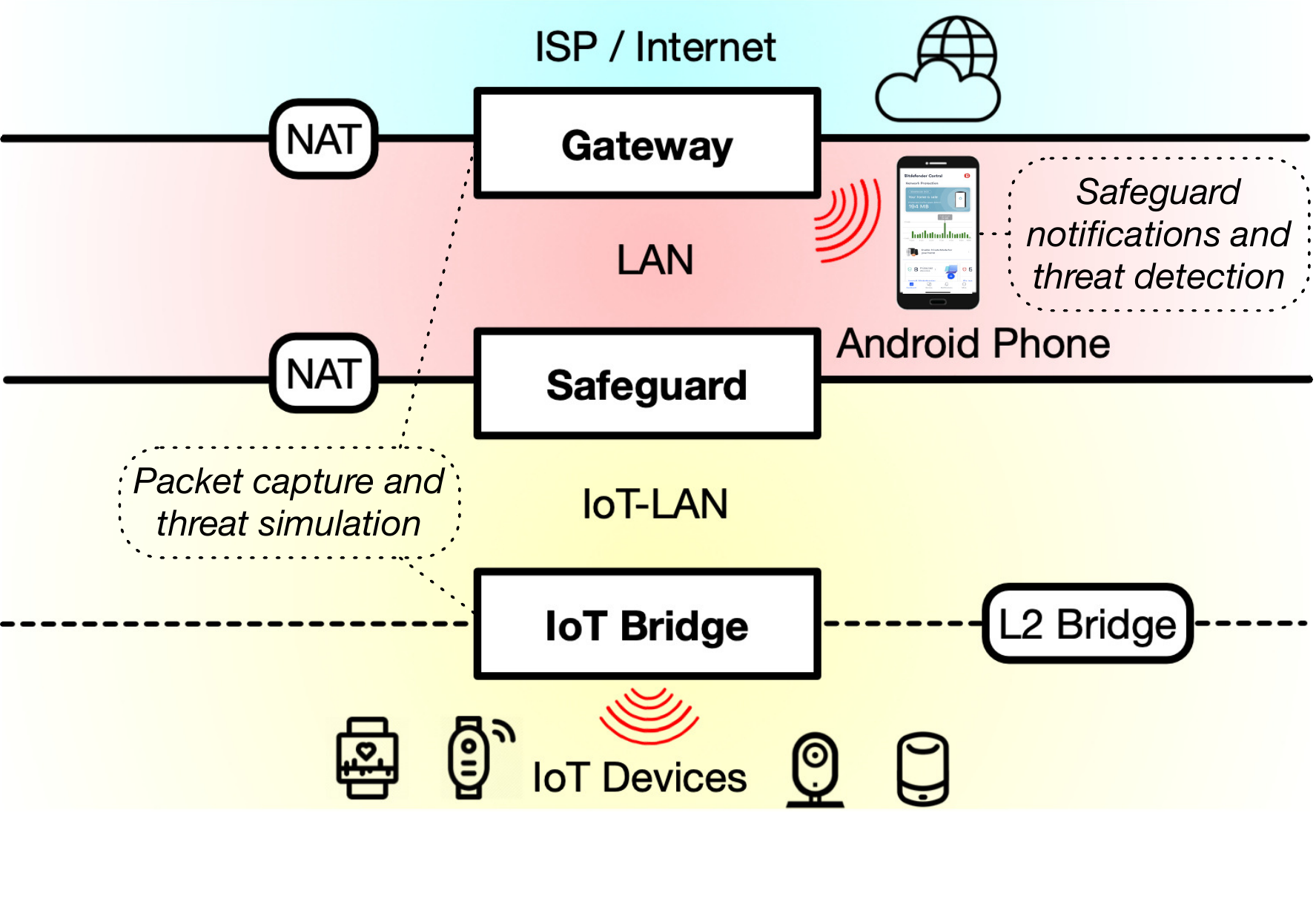}
  \vspace{-0.5cm}
  \caption{Overview of the testbed.}
  \label{fig:setup}
\end{figure}

In order to have a controlled environment for testing the safeguard, we build the testbed shown in Figure~\ref{fig:setup}. The testbed  consists of the following components: (\emph{i}) a \emph{gateway} that provides IP connectivity from the ISP to the safeguard and captures all the network traffic between the Internet and the safeguard; (\emph{ii}) the \emph{safeguard}, interposed between the gateway and the IoT-LAN in a NAT configuration, (\emph{iii}) the \emph{IoT bridge}, which acts as a Layer-2 bridge bridging together the IoT-LAN interface of the safeguard with all the IoT devices, (\emph{iv}) the \emph{IoT devices under test}, a group of popular IoT devices all connected to the IoT-LAN via the IoT bridge;
(\emph{v}) an \emph{Android phone}, connected to the LAN, with the companion apps of the safeguards installed, and controllable via the Android Debug Bridge (USB mode) at the gateway; 
(\emph{vi}) \emph{threat scripts}, run at the gateway and/or at the IoT bridge to execute IoT threat detection experiments.
More details on each testbed component are presented below.

\noindent \textbf{Gateway.}
The gateway is configured using a NAT setup. It has two network interfaces, a WAN interface with a public IPv4 address outside of any firewall, and a LAN interface with a private IP address, used to give NAT Internet connectivity to the safeguards.
The gateway manages the network on its LAN interface, providing DHCP support, assigning private IP addresses and forwarding DNS queries to the ISP DNS servers, effectively trying to match the typical configuration of a home network environment by preventing direct access to its LAN interface, while allowing the safeguards full access to the Internet.
The gateway also captures all the network traffic transiting through it using tcpdump, recording all the communication between the safeguard and the Internet.
The gateway can selectively block traffic by resolving certain DNS requests to 127.0.0.1 (instead of using the public DNS for resolution) and/or blocking certain IP addresses using simple firewall rules.
The gateway also runs \emph{threat scripts} to simulate threats originating from the Internet.
Finally, the gateway is physically connected to an Android phone via the Android Debug Bridge, and has the ability to control and retrieve data from the companion app of the safeguard.

\noindent \textbf{Safeguard.}
We assume the safeguard to be a \emph{router safeguard}, since all the other types of safeguards are also intended to be configured in a way that is equivalent to a router firewall using a user-supplied home router, as discussed in Section~\ref{sec:safeguards}. 
We connect the Internet/WAN port of the safeguard to the LAN interface of our gateway, while the other port is connected to the IoT-LAN through the IoT bridge.
The safeguard routes the traffic between the LAN and the IoT-LAN in a NAT configuration and provides full DHCP support to the IoT devices in the IoT-LAN.
Devices in the IoT-LAN also have private IP addresses, but from a different subnet with respect to the private IP addresses in the LAN network since there is a second NAT.
The safeguards are installed using default settings.
%, to mimic a normal user installation.
%This choice is to better simulare a typical home scenario where a user would put their safeguard between the ISP router and their IoT devices. \\
%In this study we have two special cases of safeguards that are \emph{transparent} and do not offer the capability to act as a router (Fingbox and RATtrap). These transparent safeguards have two network interfaces, one designed to be connected to the ISP and the other to a router where all the IoT devices are connected to. To be able to use our testbed setup with these transparent safeguards, we simply add an additional router between the second network interface of the safeguard and the IoT-LAN, while keeping the first network interface still connected to our gateway.
%After this modification the setup for routed safeguard and the setup for the transparent safeguards (i.e., safeguard without routing capabilities) are equivalent and fully comparable.
%For simplicity, we consider the additional router we add for the transparent safeguards as integral part of such safeguards.

\noindent \textbf{IoT Bridge.}
This component is a Layer-2 bridge managed by us, with two network interfaces, the first one connected to the IoT-LAN interface of the safeguard, and the other interface (Wi-Fi) connected to all the IoT devices under test.
The IoT bridge the network traffic being bridged between the IoT devices and the safeguard using tcpdump.
%Each device's traffic is filtered by MAC address into separate files.
Moreover, similar to the gateway, the IoT bridge is responsible for running certain \emph{threat scripts} to simulate threats originating from the IoT devices.
Due to the privileged role as a Layer-2 bridge, we use the IoT bridge to perform attacks that require spoofing, for example by producing network traffic on behalf of other IoT devices in the same network by spoofing their IP addresses and MAC addresses.

\noindent \textbf{IoT Devices.}
\label{sec:iot}
We consider consumer IoT devices typically deployed in a smart home. We selected these devices to provide diversity within different categories and among the most popular ones we could find on the market. We chose devices in 5 categories: Camera (12), Home automation and appliances (39), Smart hubs (10), Smart speakers (13), Video (5), for a total of 79 IoT devices. 
%To better represent how IoT devices behave in the wild, we try to keep their default configuration and privacy settings unaltered and we do not perform user- initiated firmware upgrades. The devices are still allowed to perform automated firmware upgrades when such a feature is enabled in the default configuration.
All the IoT devices are connected to the Wi-Fi interface of the IoT bridge, making them part of the IoT-LAN managed by the safeguard
(i.e., their private IP addresses and DNS servers are assigned by the DHCP server of the safeguard).
Their network traffic is captured by the IoT bridge, which also can alter their traffic to simulate threats, through the use of threat scripts.
The scale of this study is comparable to other IoT studies where real devices are used~\cite{moniotr,mandalari-pets21,kolcun2020case,saidi-imc20,huang2019iot}. All the configurable settings are left as default.
%The IoT devices and the safeguards can usually be controlled via a \emph{companion device} such as a smartphone application. 
%Our testbed allows us to perform automated experiments on the safeguards using these companion devices.
%In this case, the monitoring software captures the network traffic of both IoT and companion devices into separate PCAP files.
%The testbed allows us to capture several network traces for each device and safeguard, retry destinations and perform self-validating automated experiments under different conditions.
%Finally, the testbed allows us to spoof traffic at device level by injecting traffic to the network.

\noindent \textbf{Android Phone.}
We use an Android phone to understand if the safeguards are operating normally (i.e., they report an ``online'' or ``ready'' status), 
and to understand if and when safeguards detect threats via their companion app using threat detection scripts.
The Android phone is connected via USB to the gateway, which offers programmatic control via ADB, thus allowing the execution of threat detection scripts.
The Android phone is connected to the LAN (i.e., the network that connects the gateway to the safeguard).

%\subsubsection{Activity scripts}
%The control of IoT devices is obtained using activity scripts, using a methodology similar to~\cite{XXX}, where a function of an IoT device is triggered using a trigger activity script, and the successful execution of such function is verified using a probe activity script.
%An activity scripts execution is successful (unsuccessful) when the function of the device was correctly (incorrectly) executed.
%Similar to~\cite{XXX}, we calibrate and test activity scripts 30 times for the three main functions of all the IoT devices under test, to ensure that their accuracy is above 99\% as also obtained in previous work.

%\subsubsection{Safeguard health scripts}
%Our methodology relies on the possibility to detect if a safeguard self-reports as being online (i.e., it is working) or offline (i.e., it is not working) through its companion app.
%To detect this, we use a screenshot-based method similar to~\cite{XXX} to detect whether a safeguard is working or not.
%The control of IoT devices is obtained using activity scripts, using a methodology similar to~\cite{XXX}, where a function of an IoT device is triggered using a trigger activity script, and the successful execution of such function is verified using a probe activity script.
%An activity scripts execution is successful (unsuccessful) when the function of the device was correctly (incorrectly) executed.
%Similar to the IoT probing strategies in~\cite{XXX}, we calibrate and test safeguard health scripts 30 times for each safeguard, to ensure that they can be trusted, i.e., having accuracy above 99\%.

\noindent \textbf{Threat Scripts.}
To scale our analysis to many safeguards, we simulate threats programmatically using \emph{threat simulation scripts}, which, depending on the type of threat, are run either in the gateway (threats originating from the Internet) or on the IoT bridge (threats originating from or targeting the IoT devices under test).
Examples of simulated threats involve spoofing/replaying traffic, port scanning, etc.
The full list of threats we simulate is reported in Section~\ref{sec:rq2}. 

In addition to threat simulation scripts, we also need a way to automatically collect the detection results of a threat, i.e., if the safeguard detects a threat we simulate, or if a safeguard detects a threat we do not simulate.
To do this, we use \emph{threat detection scripts}. The threat detection scripts programmatically query the status of a safeguard by querying their control interfaces. All the safeguards we analyze offer as control interface an Android companion app and can be configured to produce an Android notification or an entry in their log files every time a threat is detected. 
Therefore, all our detection scripts leverage the safeguard Android companion apps to automatically collect information about threats that are detected (and those that are not).

%!TEX root = ../paper.tex

\section{Methodology}
\label{sec:method}

%In this section we report the methodology we use for answering our research questions.
% by proposing an experimental setup that detects the safeguards traffic and automatically classifies their properties as successful or not. 

\subsection{Characterizing Safeguard Operation}
\label{sec:rq1}

To answer our first research question, i.e., understanding the privacy and security implications of how a safeguard works, we establish a methodology to: (\emph{i}) determine if a safeguard processes its information locally or remotely; (\emph{ii}) determine which destination parties are involved in processing such information; (\emph{iii}) determine if the safeguard correctly identifies the IoT devices under test.
Accurate and timely threat detection methods need to be customized to a particular IoT device, or at least a particular service provider/manufacturer, since it is common for manufacturers to reuse infrastructure, hardware and software architectures~\cite{mandalari-pets21,saidi-imc20}. The safeguards claim to use behavior-based technology to detect and block advanced threats. In this context understanding the nature of the IoT device is essential. 

\subsubsection{Processing Locality} 

Understanding processing locality, \ie the actual location in which a safeguard processes the network traffic from its IoT devices, is important from both a privacy and security perspective.
From a privacy perspective, having IoT devices' data being processed remotely (e.g., by a cloud service) means that there is some degree of information exposure, with all its risks (e.g., data sharing, data retention, analytics, and the risk of data leaks). 
From a security perspective, having a safeguard relying on a remote service means that the safeguard will be unable to work (i.e., notify the user of possible threats) in case the remote service has an outage or is blocked, defeating the purpose of the safeguard and giving a false sense of security to its users. 
%In this section we report the methodology we use for answering our first research question. We develop a methodology for characterizing how safeguards operate, i.e., if they need the cloud to operate, if they can communicate locally or if they use third party services to operate. 
To determine whether the safeguards process the IoT data locally or remotely, we first \emph{collect the destinations} contacted by the safeguard that are not contacted by the IoT devices in the IoT-LAN, and then perform a \emph{locality experiment} to determine if the traffic to such destinations is related to the IoT devices' traffic. 

\emph{Collecting Destinations.}
We set up our testbed for one month and capture all the traffic produced by the safeguard with the IoT devices under test connected to it. The packets are captured both in the gateway (Gateway in Figure~\ref{fig:setup}) and in the IoT bridge (IoT Bridge in Figure \ref{fig:setup}).
Since the safeguard traffic captured at the gateway also includes the IoT devices' traffic (that we do not need), we filter out the IoT bridge traffic that is present in the gateway traffic to select only traffic that is generated by the safeguard.
We primarily identify all contacted destinations by hostname rather than by IP address. To do this, for each IP destination, we look at all the DNS traffic for the safeguard to find the DNS hostname that resolved to the IP address. If the hostname cannot be determined using this method, we simply use the IP address as the destination. 

\emph{Locality Experiment.}
To check whether the safeguard processes its data locally or not, we design a \emph{locality experiment}.
Specifically, we compare the IoT traffic with the traffic to the collected destinations and we verify whether the traffic is periodic or is related to the activities of the IoT devices.
We then confirm our results by looking at each safeguard's privacy policy (Section~\ref{sec:privacypolicy}).
%Specifically, we filter the list of destinations contacted by the safeguard (using DNS blocking for hostnames and firewall rules for IPs) and then look at the companion app of the safeguard (see Figure~\ref{fig:setup}) to check if it is still working (i.e., it reports an ``online'' status).
%If after filtering its destinations, the safeguard stays online, we assume the safeguard can operate locally (i.e., it uses an on-device processing strategy). If after filtering, the safeguard becomes offline, we assume the safeguard uses the cloud to operate (i.e., it offloads its processing).
%App and safeguards are under the same WiFi network provided by the router.

\subsubsection{Party Characterization}

When a safeguard offloads some of its processing to an external service, it could potentially lead to some degree of information exposure.
Given that some safeguards may not have enough on-board processing power to detect IoT threats locally, and therefore some cloud communication is inevitable, we further determine whether such communication involves more destination parties than needed.
To address this, we determine the trends related to whether a safeguard uses support or third-party destinations to operate.
We use the following definitions of parties, also used in prior work~\cite{mandalari-pets21}: a \emph{First party} is a destination related to the safeguard manufacturer; a \emph{Support party} is a destination that is not a first party and is responsible for providing remote computation or communication services (i.e., cloud computing and CDN services that are not explicitly related to advertising and analytics); and a \emph{Third party} is a destination that is neither a First party nor a Support party.
We assume that first parties and support parties are expected parties for cloud-processing devices, while we consider third parties to be likely non-essential parties, and therefore a notable source of risk of private information exposure.
Previous IoT work~\cite{moniotr,mandalari-pets21} has shown a correlation between third parties (other than cloud services and CDNs) and destinations that are non-essential for device functionality. Thus, a third party is often an unneeded party (though this may not be true for every third party).

\subsubsection{IoT Device Identification}
Device detection is challenging~\cite{kolcun2020case}, as differences among IoT devices frustrate attempts to build models that generalize to all IoT devices.
%The reason is that most IoT devices have really little in common. 
Even devices belonging to the same category (i.e., having similar sensors and similar functionality) may have totally different hardware and software architectures.
For this reason, accurate threat detection methods need to be customized to a particular IoT device, or at least a particular manufacturer, since manufacturers often reuse hardware and software architectures within the same generation of their devices~\cite{mazhar2020characterizing,saidi-imc20}.
To address this, all the safeguards include, as the basis for their operation, a feature to automatically identify the IoT devices they protect.
Our hypothesis is that, if an IoT device is correctly identified, the safeguard can apply a more efficient and specific threat detection model to protect such device, thus resulting in better security.
Conversely, if the IoT device is not correctly identified, it may apply the model for the wrong device, or apply a generic model that is not tailored for that specific IoT device. 
Five safeguards offer ``Advanced Threat defense,'' a behavior-based technology to detect advanced threats and ransomware. Unfortunately they do not reveal any details; however, we do learn that they implement some device-specific features. For example, Bitdefender allows the user to ``mute'' smart speakers; this feature can be applied to smart speakers only. This function disables the smart speaker's ability to eavesdrop on conversations by temporarily severing their connection to the Internet. This highlights the importance of correctly identifying the IoT device. 

%All the safeguards under test claim to automatically discover and identify all devices that connect to the same network.
To test the IoT device detection capabilities of our safeguards, we iterate the following three steps 10 times: (Step 1) we reset the safeguard to its factory settings and connect it to our testbed while all our 79 IoT devices are disconnected; (Step 2) we connect the 79 IoT devices and wait for 30 minutes; (Step 3) we check the IoT devices identified in the safeguard companion app.
We classify each IoT device as: \emph{detected} if the IoT device is correctly identified; \emph{mislabeled} if the IoT device is identified as something else; \emph{unknown} if the IoT device is reported as ``unknown'' or it has no label.

\subsection{Threat Detection Capability}
\label{sec:rq2}

This section describes the methods we use for answering our second research question, \ie investigating the capability of the safeguard to identify the threats they claim to detect.
The threat detection capabilities we test are against (\emph{i}) \emph{security threats}, spanning specific attacks (e.g., flooding) to generic anomalies (\eg a device behaving in an unexpected way), and (\emph{ii}) \emph{privacy threats}, spanning the presence of unencrypted traffic to the proper enforcement of DNS over HTTPS (DoH)~\cite{bumanglag2020impact}. 
\new{We test a representative sample of already-exploited attacks on IoT that safeguards claim to detect. Although not all such threats are ``IoT-specific'', they still affect IoT devices, and we therefore include them in our study to ensure that commercial IoT safeguards fulfill their promises.}
In the remainder of this section we describe the detection capabilities and the threat scripts we use to simulate/detect such threats.
%Note that some safeguards are also able to detect suspicious network traffic (e.g., anomalous traffic patterns), potential vulnerabilities (e.g., open ports), and to perform mitigation actions (e.g., device quarantining). 
%We also include such cases in the list below, since although they are not technically attacks, they are closely related to them.
%Our methodology is designed to systematically test the attack depection  the safeguards declare to have. 
%We describe all specialized device triggers and probes we use for each of the feature.

\subsubsection{List of Detectable Threats}
%!TEX root = ../paper.tex

\begin{table}[t!]
\centering
	\caption{List of threats and threat simulation scripts}
	\vspace{-2mm}
	\label{tab:method}
	\resizebox{0.85\columnwidth}{!}{
		\centering
		
		\begin{tabular}{m{0.001\columnwidth}p{0.3\columnwidth}||c}
		\centering

			&\textbf{Threat} & \textbf{Threat Simulation Script}  \\ 
			\hline
			\hline
%Security
\multirow{14}{*}{\rotatebox[origin=c]{90}{\scriptsize Security}} 
& Anomalous ON/OFF & Power Script \\
\cline{2-3}
& Anomalous Traffic & Anomalous Traffic Script \\
\cline{2-3}
& Anomalous Upload & Anomalous Upload Script  \\
\cline{2-3}
& Open Port  & Kali-based Script  \\
\cline{2-3}
& Weak Password & FTP-based Script  \\
\cline{2-3}
& Device Quarantine & None (manual via app) \\
\cline{2-3}
& SYN Flooding  & Kali-based Script  \\
\cline{2-3}
& UDP Flooding  & Kali-based Script  \\
\cline{2-3}
& DNS Flooding & Kali-based Script \\
\cline{2-3}
& HTTP Flooding & Kali-based Script  \\
\cline{2-3}
& IP Fragmented Flooding & Kali-based Script  \\
\cline{2-3}
& Port Scanning & Nmap-based Script  \\
\cline{2-3}
& OS Scanning & Nmap-based Script \\
\cline{2-3}
& Malicious Destinations & Malicious Destinations Script  \\
\hline
\hline
%Privacy
\multirow{3}{*}{\rotatebox[origin=c]{90}{\scriptsize Privacy}} 
&  PII Exposure & Privacy Script \\
\cline{2-3}
&  Unencrypted Traffic & Privacy Script \\
\cline{2-3}
&  DoH & DNS Script \\
\hline
\hline

		\end{tabular}
	}
\end{table}

In this section we describe the threats we consider in our tests (see the first column of Table~\ref{tab:method} (Threat)).
To be included in our list, a threat must be declared as detectable by at least three safeguards. 
The reason for this is that most safeguards do not publicly disclose what threats they really can detect. Our inference from the testbed evaluations is that if some safeguards detect them, then others might too.
Note that we reset the safeguard to its factory settings for the IoT device identification test only, not for the threat detection.

\noindent \emph{Anomalous ON/OFF}. Safeguards claim to detect anomalous behavior, assuming that it might be the result of a security attack. To test this, we emulate the situation in which a compromised device is switched on and off continuously. 
%Safeguards claim to detect when a device is doing \emph{strange} things (i.e, it exchanges more data than it would be normal with the outside world). This anomaly emulate a compromised devices that is switched on and off continuously. 

\noindent \emph{Anomalous Traffic Patterns}. Safeguards claim to use Intrusion Detection System (IDS) and Intrusion Prevention System (IPS) technology to detect if an IoT device is producing unusual traffic, \eg as the result of an attack. To test this, we produce synthetic traffic that does not conform to the typical traffic of an IoT device. Some safeguards claim to use \emph{behavior-based technology}; we do not know if the safeguards are designed to ``learn'' normal traffic or if they use some pre-trained models. Under cover behavior-based detection, we connect the IoT devices to the safeguards for one month, and assume this is enough for a safeguard to ``learn" IoT device behavior.

% can understand the intent of the attacker (or user), and based on the intent, it can take action, generate alarms, or block. Some safeguards claim to be able to detect unusual upload or transfer of data. They refer to this as anomaly-based detection.

\noindent \emph{Anomalous Upload}. Safeguards claim to detect when some IoT devices generate anomalous bursts of outgoing traffic. 
%Anomalous uploads are types of anomalies that are notifying the users when a peak of traffic is detected. Safeguards study how IoT devices upload/transfer to the internet.  
For example, when a user remotely connects to a smart camera, the camera will upload a video stream. However, if this behavior is not triggered by a legitimate user or by a camera device, the safeguards should detect and report the anomalous upload as a potential threat.
% this activity and classify it as 'abnormal'.  

\noindent \emph{Open Port}. 
While this is not always a threat, it is a potential threat detected by some safeguards:
attackers commonly use port scanning software to find which TCP ports are ``open'' (unfiltered) in a given device, and whether or not a service is listening on that port, with the goal of exploiting security vulnerabilities to gain unauthorized access.
A safeguard supporting this detection capability must notify the user when specific ports are open on an IoT device.

\noindent \emph{Weak Password}. The safeguards claim to notify the user when weak passwords are used. % and asses this vulnerability.

\noindent \emph{Device Quarantine}. This is not technically a threat \emph{per se}, but a threat mitigation strategy offered by some safeguards: when a threat is detected from an IoT device, the device should be isolated, \ie put in quarantine. We test this by simulating a detectable attack and checking whether the IoT device was properly quarantined.

\noindent \emph{SYN, UDP, DNS, HTTP, IP Fragmented Flooding}. These are the five most common classes of DoS attacks a Mirai-infected device will run~\cite{antonakakis2017understanding}. The safeguards that use machine learning and signature-based algorithms claim they can detect and block such attacks.

\noindent \emph{Port Scanning}. A port scan attack is used to find open ports and figure out whether they are receiving or sending data.

\noindent \emph{OS Scanning}. An OS scan is an attack for retrieving information about the OS (and version) of a device.

\noindent \emph{Malicious Destinations}. A compromised IoT device on the home network may connect to malicious destinations. Some safeguards claim to detect traffic to well known malicious destinations. We test this by simulating traffic to popular malicious destinations (i.e., \url{https://easylist.to/}).

\noindent \emph{PII Exposure}. Sensitive personally identifiable information (PII) should be transmitted and stored securely (\eg using encryption). 
Some safeguards claim to detect when PII are exposed in plaintext.

\noindent \emph{Unencrypted Traffic}. According to best practices and some privacy regulations~\cite{wachter2018normative}, network traffic should be sent encrypted. 
Some safeguards declare to protect the privacy of the user by checking that the device uses encryption. 

\noindent \emph{DoH}. DNS over HTTPS (DoH)~\cite{rfc8484} is a protocol for performing remote DNS resolution via the HTTPS protocol. 
DoH can improve privacy by preventing ISPs from seeing DNS requests in plaintext. 
Some safeguards allow the user to use DoH. 
If DNS over HTTPS is enabled on some safeguards, the IoT device will be ``network-patched'' to use the DoH server even if it has its own DNS server configured.

\subsubsection{Threat Simulation Scripts}
\label{sub:triggers}

As discussed in Section~\ref{sec:testbed}, we use threat simulation scripts that are specific to each simulated threat to reproduce the threats that the safeguards claim to detect.
The scripts we use for each detectable threat is reported in the second column of Table~\ref{tab:method} (Threat Simulation) and described as follows.

\noindent \emph{Power Script.}
This threat simulation script is used for emulating the anomalous ON/OFF behavior. 
All IoT devices are plugged into remotely controllable smart plugs, and we use power scripts to turn these smart plugs on and off so that we can reset the IoT devices by power-cycling them.
We use the power script for emulating a situation in which the devices are compromised and are switched on and off continuously. 
The devices are power-cycled every 30 seconds for one hour.

\noindent \emph{Anomalous Traffic Script.}
This script emulates an anomalous traffic pattern by emulating Google Home traffic as being produced by an Echo Spot.
First, we collect normal traffic produced by an Echo Spot while interacting with the device.
%We also collect traffic from another IoT device (Google Home) and record pcap files, logging all packets sent during that time period. We performed many interactions that would occur during regular device use, 
 % to the smart speakers. 
From the IoT bridge we use  tcpreplay~\cite{tcpreplay} spoof source IP address, MAC address, and inject the Echo Spot traffic to the safeguard as if the Echo Spot is sending traffic as Google Home. 
\new{We assume that substantially changing the nature of traffic from an IoT device would be classified as anomalous behavior under reasonable definitions. Swapping the traffic of two devices results in different destinations, different protocols, and different traffic patterns. 
In fact, multiple safeguard vendors confirmed that the safeguards learn the behavior of the IoT device using network traffic for $n$ days of learning and may include previous knowledge from the device  (see Section~\ref{sec:discussion}). We therefore assume that a sudden change in traffic behavior on the device should be detectable by the safeguards.}

\noindent \emph{Anomalous Upload Script.}
To emulate this behavior we instruct the IoT bridge to simulate the traffic patterns of a smart camera that is uploading a video, then we spoof its IP address, MAC address, and packet timings to make it appear as if all the IoT devices (including non-cameras) simultaneously produced camera traffic using tcpreplay~\cite{tcpreplay}.

\noindent \emph{Kali-based Scripts.}
Kali Linux contains a suite of attacks we leverage as threat simulation scripts. 
We use Kali to simulate the five most common classes of attacks a Mirai-infected device will run~\cite{antonakakis2017understanding}. 
We run a Kali Linux  virtual machine running on the IoT bridge (as the attack source when the threat is originating from the IoT devices) or the gateway (as the attack source when the threat is originating from the Internet) and two Web Servers (as attack victims). 
We spoof source IP address, MAC address, and inject the traffic produced by Kali attacks to the safeguard to make it appear as if the Echo Spot is infected and is sending DoS traffic in addition to its normal traffic. 
\new{For weak passwords, we use a script to generate traffic using FTP that exposes the safeguards to plaintext protocol sessions containing 200 common weak passwords.  Vendors confirm they perform this kind of scanning (Section~\ref{sec:discussion}).}

\noindent \emph{Nmap-based Script.}
Nmap~\cite{lyon2008nmap} is an open source utility for network discovery and security auditing. 
We use a threat simulation script that uses Nmap on the IoT bridge and the gateway to launch port and OS scan attacks.

\noindent \emph{Malicious Destinations Script.}
The malicious destinations script simulates a device contacting malicious destinations: it
instructs the IoT bridge to spoof the Google Home device, and to contact, while spoofed, 83,021 malicious destinations, of which 39,354 are verified as active.
%We configure the IoT bridge so that it spoofes the Google Home device. We then contact 83021 malicious destinations, of which 39354 are active.
%We first initiate a connection to the destinations with no safeguards and exclude the non-responsive destinations. 

\noindent \emph{Privacy Script.}
We design a script for emulating privacy leakage. 
We use the IoT bridge capability of spoofing the Google Home device and inserting, using tcpreplay, some plaintext traffic that includes PII.
\new{We consider the PII of the account we set up for the safeguard (i.e., email, name, password, etc.)}. 
We use the same script also for verifying if a safeguard can detect unencrypted traffic.

\noindent \emph{DNS Script.}
To test the capability of a safeguard to detect and/or make DNS requests DoH compliant, we use a threat script that triggers DNS queries from an IoT device that is not DoH compliant and detects if such requests are still not compliant after being processed by the safeguard. 
%, then tracked by at the gateway level.
%We design a script for tracking all the DNS queries from one of the IoT device connected to the safeguards.

\subsubsection{Threat Detection Scripts}
%\label{sub:probes}
%
%We use one strategy for probing the devices;
%the probing strategies we use for each device are reported in the third column of Table~\ref{table:probe} and described as follows.
%
%\noindent \textbf{Companion App.}
We create a threat detection script for each safeguard, to understand the detection capabilities of a safeguard with respect to the simulated threats.
Each threat detection script is run at the gateway, and it uses the Android Debug Bridge~\cite{adb} to access the safeguard companion app, previously installed on the Android phone of our testbed.
After each threat simulation script execution is complete, the threat detection script checks the notifications and logs from the safeguard app, 
and records each threat detected by the safeguard, along with the timestamp.
\new{Since we treat the safeguards as black boxes and we rely on what they report, it is not possible for us to know what a safeguard detects, if it does not report it. 
For this reason, we consider only reported threats as detected.}
%Also, from a user perspective there is no difference between a ``no detection" and a ``detection without alert".

%We install this app on an Android phone that is \emph{not on the same LAN} as the Safeguards (to force the communication to happen over the Internet rather than directly).
%For the majority of IoT devices, a companion app is available as a method to obtain their state.
%
%For this method, we use the Android Debug Bridge\cite{adb} for checking the notification from the safeguard app, each time an event has been detected. We then log the  event notified and the time of the event in the router. 

 \begin{figure}
  \centering
  \includegraphics[width=0.85\linewidth]{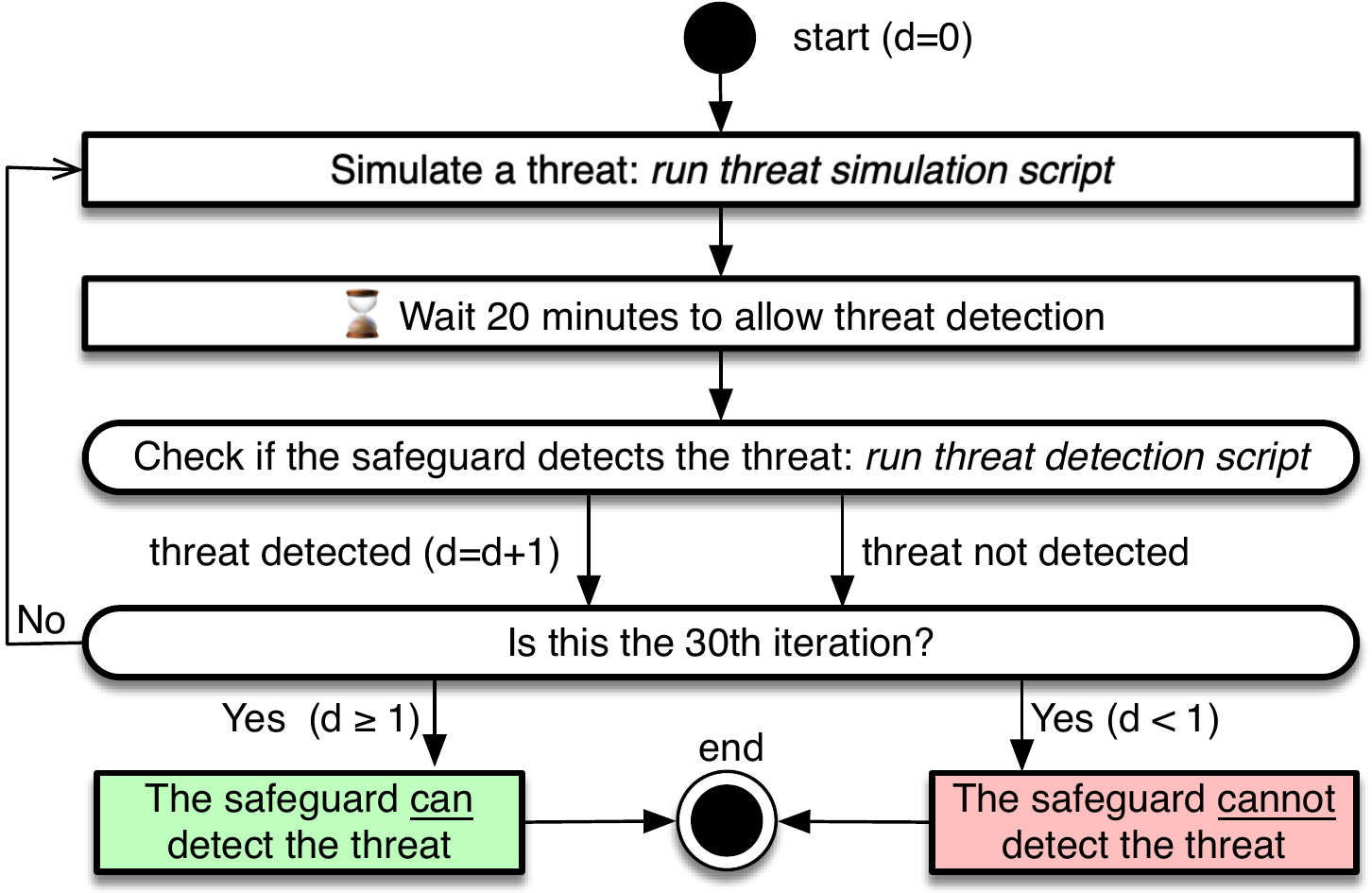}
    \vspace{-0.15cm}
  \caption{Threat detection capability experiment.}
  %We simulate a threat 30 times for each safeguard: if the threat is detected by the safeguard at least once, we consider the detection capability satisfied.
  \label{fig:experiment}
\end{figure}

\subsubsection{Threat Detection Experiments}

We use threat detection experiments to first simulate a threat, and then check whether such threat has been detected or not.
Each threat detection experiment iterates the following for 30 times (see Figure~\ref{fig:experiment}):

%
%The basic unit of our methodology is the \emph{feature experiment} (see Figure~\ref{fig:experiment}).
%We define it as the fully automated process to verify if a feature of a safeguard can be executed or not.
%
%To ensure that we can reliably use information from probe scripts that may be inaccurate, we test the function \emph{multiple times} (at least 30), and we consider the result correct if a strong majority (at least 80\%) of the test results are consistent.
%This ensures that the whole functionality experiment has a negligible probability of an incorrect result:
%while the odds of any one test failing can be significant, the odds that all of the multiple tests fail is substantially lower.
%
%Specifically, each feature experiment iterates at least 30 times the following three steps. 
%

\noindent \emph{Step 1:} We execute the threat simulation script corresponding to the threat detection capability we want to test.

\noindent \emph{Step 2:} We wait a fixed amount of time to give the safeguard enough time to detect the threat. 
%The wait time is determined empirically as an amount of time that is longer than the longest amount of time it takes for a safeguard to detect a threat. 
We have empirically found that 20 minutes are enough since the longest time to detect a threat among all safeguards has been 3 minutes.

\noindent \emph{Step 3:} We execute the threat detection script to determine whether the threat was detected by the safeguard under test. 
If the threat is detected in at least one iteration, we consider the detection capability satisfied.

\new{
While our experiments are automated via threat simulation and detection scripts, the creation of such scripts and the data they use (e.g., network traffic) is a manual process that is safeguard- and threat-dependent.
However, once scripts are written, they can be reused across experiments and need to be rewritten only after major changes in the safeguard interaction interface.
We found that manual tests were seldom necessary for features that could not be automated, \eg testing device quarantine (which was triggered manually in safeguards apps).
%and open port detection (port scan manually triggered in safeguards apps).
}
\subsection{Safeguard Side Effects}
\label{sec:rq3}

To answer the last research question, i.e., detecting what side effects a safeguard causes that can affect their adoption, we perform three additional investigations: a performance analysis focused on network overhead, a false positive analysis to assess user fatigue, and finally we analyze the presence and the content of the privacy policy of the safeguards. We connect the IoT devices to the safeguards for one month.

\subsubsection{Overprotection} 
Some safeguards could overprotect and mislabel some traffic as a threat and notify the user. 
To verify this, we connect 12 IoT devices to the safeguards and we capture the traffic on the router for one month. 
We use the threat detection scripts to log the notifications received from the safeguard in that period. 
If we get a notification, we then inspect the packet capture to verify if it was a false positive or not. 
A large number of false positives may weaken users' trust in detected threats, and lead to alert fatigue 
that can cause users to ignore alerts or disable the safeguard. % entirely.

\subsubsection{Network Traffic Overhead}
\label{sec:rq3_traffic}
To understand the impact of the safeguard on IoT-device performance, we measure (for one month) what percentage of traffic is network overhead caused by the safeguard. To do so we capture the traffic at the gateway $T_G$, which is the sum of the traffic produced by the safeguard ($T_S$) and the traffic produced by the IoT devices connected to it ($T_D$).
Since the gateway cannot distinguish $T_S$ from $T_D$ because of the NAT configuration, we measure $T_D$ using the local IoT traffic captured by the IoT bridge.
Finally, we calculate the safeguard traffic as $T_S=T_G-T_D$ and compute the network traffic overhead $\textit{ov}$ as the fraction of network traffic produced by the safeguard over the total traffic: $\textit{ov}=\frac{T_S}{T_G}$.
A large overhead value might impact the user experience, which could lead users to disable the safeguard to improve performance at the expense of security and privacy.
%(the IoT bridge shares the same network with all the IoT device and is therefore able to separate the traffic for each device from the safeguard traffic using MAC address). 
%Finally, now that we have  we then compute the percentage of Bandwidth overhead from each safeguard. 

\subsubsection{Privacy policy} 
All the safeguards need access to the network traffic produced by the IoT devices; however, 
the way that data is processed, used, and shared over the Internet can have a significant impact 
on privacy and security.   
%In case of devices that rely on the cloud, this traffic (or some information inferred from the traffic) may be sent to the cloud for remote analysis.
In fact, both to be compliant with privacy regulations and to address user concerns, each safeguard must declare what data they collect, for which purpose, for how long, and if it is being shared and why.
To shed light on these issues, we manually analyzed the privacy policy of each safeguard to understand their data use and sharing disclosures.
We expect them to use their data in anonymized form, to retain it only for the minimum amount of time necessary to perform the analysis and the detection, and not to share it with third parties; however, as we will show in Section~\ref{sec:privacypolicy}, not every safeguard makes this clear, thus creating concerns for their users, and possibly limiting the adoption of their solution.

\subsection{Validation}
Our methodology allows us to verify that the threats go through the safeguards and reach the IoT devices (and vice versa). Packet capture happens at the \emph{Gateway} and at the \emph{IoT Bridge}. For all experiments we verify that the threats correctly traverse the safeguards. We validate the threats we produced using \emph{Snort}~\cite{snort_link,roesch1999snort,gupta2019mitigation}. Snort is a powerful open source network intrusion detection system, capable of performing real time traffic analysis and packet logging on IP networks. All the threats we produced have been successfully detected by Snort. Moreover, when performing the malicious destinations experiment, our institution received an alert from our National Cyber Security Centre advising us of suspicious network activity originating from a device that we have recently used. This further demonstrates the visibility and validity of our experiments.

%!TEX root = ../paper.tex
\section{Evaluation of Safeguard Operation}
\label{sec:rq1_evaluation}

%!TEX root = ../paper.tex

\begin{table}[t!]
\centering
	\caption{Summary of the strategy of the safeguards}
	\label{tab:strategy}
	\vspace{-2mm}
	\resizebox{1.0\columnwidth}{!}{
		\centering
		\begin{tabular}{p{0.07\textwidth}||c|c|p{0.23\textwidth}}
		\centering
			\textbf{Safeguard} & \textbf{Destinations \#} & \textbf{Cloud} & \textbf{\# and list of Support/3rd Parties}\\ 
			\hline
			\hline
%General
Avira &  10 &Yes & (1) api.mixpanel.com     \\
\cline{1-4}
Bitdefender &  5 &Yes & - \\
\cline{1-4}
F-secure &  1 &Yes & -   \\
\cline{1-4}
FingBox &  5 &Yes & (2) api.snapcraft.io, 

mlab-ns.appspot.com   \\
\cline{1-4}
Firewalla &  4 &No & (1) api.github.com   \\
\cline{1-4}
McAfee &  22 &Yes & (3) app-measurement.com, 

commscope.com, avast.com   \\
\cline{1-4}
RatTrap &  1 &Yes & -   \\
\cline{1-4}
TrendMicro &  3 &Yes & (1) policy.ccs.mcafee.com   \\

\hline
\hline	
		\end{tabular}
	}

\end{table}

We now answer our first research question by using the methods from Section~\ref{sec:rq1} to identify and characterize security failures for our safeguards. 
%In particular, we build, for each safeguard, the list of destinations contacted, and then perform analyses to 
In particular, we identify how safeguards operate, locally or using the cloud, and if they correctly identify the devices they are protecting. 

\subsection{Processing Locality and Party Analysis}
\label{sec:strategy}

We now characterize the destinations of the traffic of each safeguard.
Table~\ref{tab:strategy} shows that 7 safeguards process the IoT data remotely, while one of them (Firewalla) works locally.  
F-secure and RATtrap rely on 1 destination to correctly work, while the rest of the safeguards contact more destinations. 
Avira, Fingbox, Firewalla, McAfee and TrendMicro contact at least one destination that is not first party. 

Avira contacts Mixpanel (\url{api.mixpanel.com}),  a business analytics service~\cite{mixpanel}. Avira, in its privacy policy, states this destination is contacted, but that the IP address of the user is anonymized; however, the safeguard contacts this destination directly, so Mixpanel may potentially learn and store the IP address of the user, which can further be used for IP geolocation (see Section~\ref{sec:privacypolicy}).
McAfee contacts Firebase (\url{app-measurement.com}), an app analytics tool offered by Google.
Fingbox relies on third party services that include Snapcraft and M-Lab. These destinations are best explained by the fact that the safeguard provides a speed test service to the user. Such communication can be a privacy concern, \eg if these third parties store/share IP addresses and/or other PII.
TrendMicro contacts a destination owned by McAfee---a surprising result since such safeguards do not have any publicly declared business agreement.

\noindent \textbf{Takeaways}: The majority of the safeguards use the cloud for performing their analysis, potentially leaving the user vulnerable in the event of a data breach. Moreover, we observe several destinations contacted that are not first parties, potentially leading to privacy and security risks.

\subsection{IoT Device Identification}
\label{sec:identification}

 \begin{figure}[t!]
  \centering
  \includegraphics[width=0.85\columnwidth]{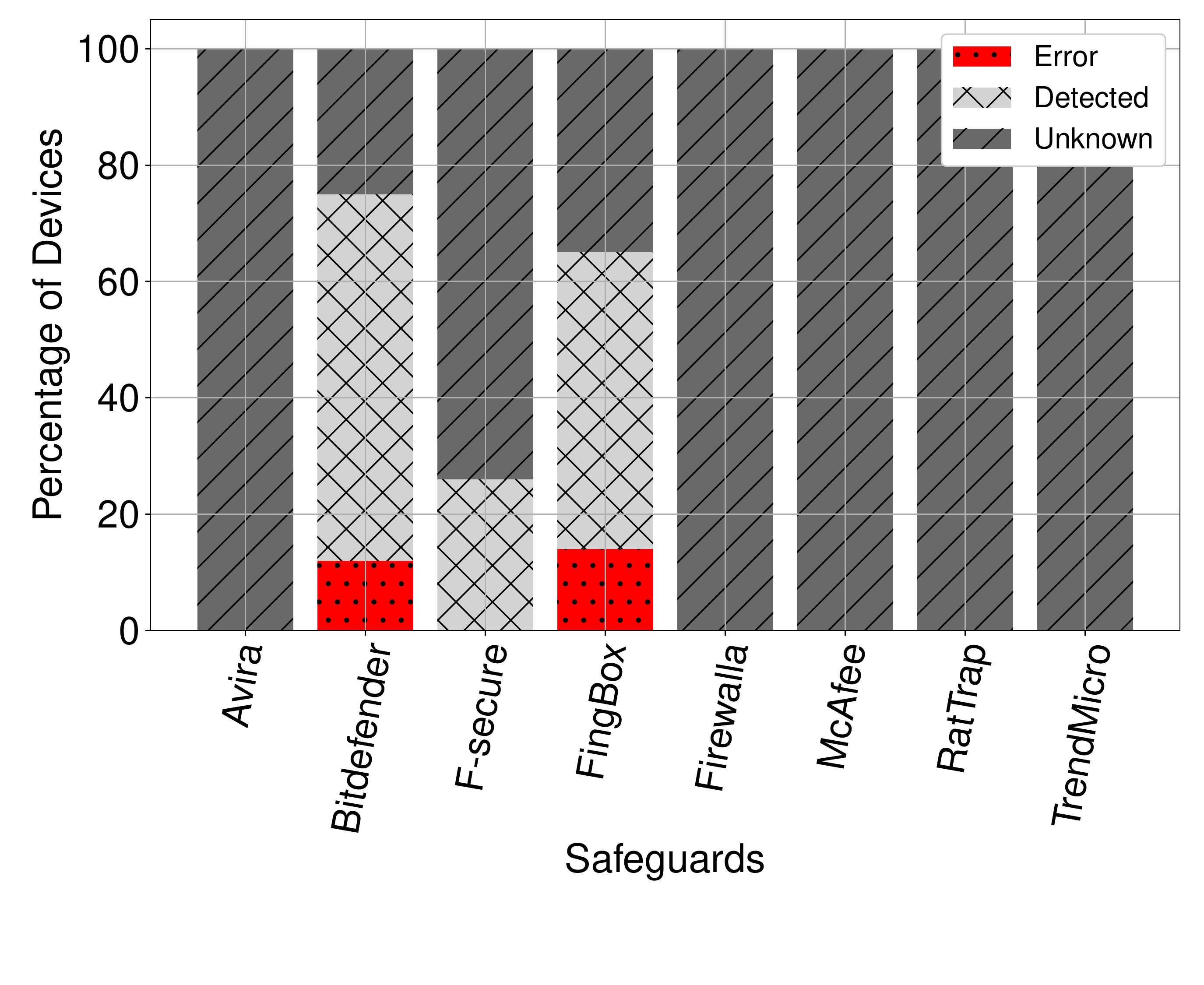}
  \vspace{-0.4cm}
  \caption{Percentage of devices correctly identified by the safeguard (Detected), mislabeled (Error), and not identified (Unknown).}
  \label{fig:identification}
\end{figure}

In this section, we determine trends relating to whether the safeguard is able to correctly identify the IoT devices connected to it. We connect 79 IoT devices to the safeguards; Figure~\ref{fig:identification} shows the result of the identification per safeguard. We use the following definitions: \emph{Detected} is the percentage of IoT devices correctly identified by the safeguard; \emph{Error} is the percentage of IoT devices mislabeled (i.e., Echo Spot identified as Google Home); \emph{Unknown} is the percentage of unspecified IoT devices reported by the safeguard.
Only 3 safeguards identify at least one IoT device connected to them. 
Fingbox uses the content of the ARP packet to report the name of the IoT device. Bitdefender and F-secure use a more sophisticated methodology and report a more user-friendly name, such as \emph{Amazon Echo Speakers}.

\noindent \textbf{Takeaways}: A key finding is that, for the IoT devices we connect to the safeguard, only a small percentage is correctly identified; the rest of the devices require manual labeling. This introduces a security and privacy risk, since some safeguard features, like for example \emph{mute all smart speakers}, or protection techniques applied to specific vendors, may not work if the device is not correctly identified. 
%!TEX root = ../paper.tex
\section{Evaluation of Threat Detection Capability}
\label{sec:rq2_evaluation}

%!TEX root = ../paper.tex

\begin{table*}[t!]
\centering
	\caption{Evaluation of the threats. {\cmark} indicates whether the safeguard correctly detects the threat or not ({\xmark}) and the detection time. - indicates that the safeguard does not claim to detect the threat.}
	\label{tab:evaluation}
	\vspace{-2mm}
	\resizebox{0.8\textwidth}{!}{
		\centering
		\begin{tabular}{m{0.01\textwidth}p{0.166\textwidth}||c|c|c|c|c|c|c|c}
		\centering
%			\textbf{Device} & \textbf{\#} & \textbf{\#} & \textbf{Non-required}\\ 
			&\textbf{Threat} & \textbf{Avira} & \textbf{Bitdefender} & \textbf{F-Secure}& \textbf{Fingbox}& \textbf{Firewalla}& \textbf{McAfee}& \textbf{RATtrap}& \textbf{TrendMicro}\\ 
			\hline
			\hline
%General
\multirow{14}{*}{\rotatebox[origin=c]{90}{\scriptsize Security}} 
& \hspace{1mm} Anomaly ON/OFF & \centering - & \xmark& \xmark& -& \xmark& \xmark& \xmark& -   \\
\cline{2-10}
&\hspace{1mm} Anomaly Traffic Pattern & \centering - & \xmark& \xmark& -& \xmark& \xmark& \xmark& -    \\
\cline{2-10}
&\hspace{1mm} Abnormal Upload & \centering - & \xmark& \xmark& -& \xmark& \xmark& \xmark& -    \\
\cline{2-10}
&\hspace{1mm}  Open Port & \centering\xmark & \cmark(30s)& -& \xmark& \cmark(30s)& \xmark&-& \xmark   \\
\cline{2-10}
&\hspace{1mm}  Weak Password & \centering\xmark & \xmark& -& -& -& \xmark&-& \xmark  \\
\cline{2-10}
&\hspace{1mm}  Device Quarantine & \centering - &  \cmark& -& \cmark& \cmark& -& \xmark& -    \\
\cline{2-10}
& \hspace{1mm} SYN Flooding & \centering\xmark & \cmark(30s)& \xmark& -& \cmark(40s)& \xmark& \xmark& \xmark \\
\cline{2-10}
& \hspace{1mm} UDP Flooding & \centering\xmark & \xmark& \xmark& - & \xmark& \xmark& \xmark& \xmark   \\
\cline{2-10}
& \hspace{1mm} DNS Flooding & \centering\xmark & \xmark& \xmark& -& \xmark& \xmark& \xmark& \xmark \\
\cline{2-10}
& \hspace{1mm} HTTP Flooding& \centering\xmark & \cmark(3m)& \xmark& -& \cmark(2m)& \xmark& \xmark& \xmark   \\
\cline{2-10}
& \hspace{1mm} IP Fragmented Flooding& \centering\xmark & \xmark& \xmark& -& \xmark &\xmark& \xmark& \xmark  \\
\cline{2-10}
%& \hspace{1mm} Brute Force Attack &&&&&&& \\
%\cline{2-10}
& \hspace{1mm} Port Scanning& \centering\cmark(45s) & \xmark& \xmark& -& \xmark& -& \xmark& \cmark(30s)    \\
\cline{2-10}
& \hspace{1mm} OS Scanning& \centering\cmark(45s) & \xmark& \xmark& -& \xmark& -& \xmark& \xmark   \\
\cline{2-10}
& \hspace{1mm} Malicious Destinations& \centering\cmark & \cmark& \xmark& -& \cmark& \xmark& \xmark& \cmark   \\
\hline
\hline
%Privacy	
\multirow{3}{*}{\rotatebox[origin=c]{90}{\scriptsize Privacy}} & \hspace{1mm} PII Exposure  &\centering\xmark&\xmark& -&-&\xmark&-&-&-   \\
\cline{2-10}
& \hspace{1mm} Unencrypted Traffic &\centering\xmark&\xmark& -&-&\xmark&-&-&-   \\
\cline{2-10}
& \hspace{1mm} DoH  &\centering\xmark&\cmark& -&-&\cmark&-&-&-   \\
\hline
\hline
		\end{tabular}
	}

\end{table*}

In this section we answer our second research question by 
applying the methodology in Section~\ref{sec:rq2} to 
identify and characterize the threat detection capability of the safeguards. 
%Table~\ref{tab:evaluation} shows the results of the threats we test. The list and the details of the threats is reported in Section~\ref{sec:rq2}.

\subsection{Security Threats} 
We now characterize the behavior of the safeguards for the security threats (Table~\ref{tab:evaluation} Security). Of these safeguards, Bitdefender and Firewalla cover the largest number of security threats (35\% of threats detected).

\noindent \emph{Anomalous Threats.} 5 out of 8 safeguards claim to implement anomaly detection, including inspecting traffic pattern and detecting abnormal upload. None of the safeguards tested detect such threats.

\noindent \emph{Open port.} Only 2 out of 6 safeguards claiming to detect this threat actually identify it: Bitdefender and Firewalla. For these two safeguards we also measure the detection time, by computing the time between starting the threat simulation and receiving the notification from the safeguard. 
This threat is detected in 30-40 seconds.
Notably, Avira, Fingbox, McAfee and TrendMicro do not detect this threat.

\noindent \emph{Weak Password.} None of the 4 safeguards claiming to detect this threat actually detect it.

\noindent \emph{Device Quarantine.} 4 safeguards claim to be able to to detect this threat. Only 3 of them actually quarantine the IoT device when needed (Bitdefender, Fingbox, Firewalla).  

\noindent \emph{Flooding Attacks.} Only 2 safeguards detect SYN and HTTP flooding attacks (Bitdefender and Firewalla). F-Secure, McAfee, RATtrap and TrendMicro do not detect any attacks including a simple SYN flooding---a surprising result since such safeguards advertise this capability. 
For the two safeguards detecting SYN and HTTP Flooding, we measure the detection time as the difference between the start time of the threat and the time of the detection notification. 
SYN Flooding is detected in 30--40 seconds, while it takes 2--3 minutes to detect a HTTP Flooding.
Table~\ref{tab:evaluation} shows that no safeguards detect UDP-based attacks, such as UDP Flooding, DNS Flooding, and IP Fragmented Flooding---regardless of whether the source of the threat originates from the IoT device or the Internet.

\noindent \emph{Scanning Attacks.} The only safeguard that successfully detects port scanning and OS scanning is Avira, notifying the user after 45 seconds from when the attack starts. 
TrendMicro detects port scanning in 30 seconds, but it does not detect OS scanning. 
Surprisingly, the rest of the safeguards do not detect any scanning attacks. 

\noindent \emph{Malicious Destinations.} This is the only threat detected by more than three safeguards. When a malicious destination is contacted, Avira, Bitdefender, Firewalla and TrendMicro notify the user immediately. An explanation for this is that the malicious destinations we use are well known, and therefore easier to detect. Bitdefender is likely using a semi-supervised machine-learning approach to detect malicious destinations, as stated in one of their papers~\cite{gabriel2016detecting}.

\noindent \textbf{Takeaways}: In general, we observe that the number of undetected threats tends to be larger than the number of the detected ones---on average, only 3 out of 14 threats are detected by the safeguards. 
3 out of 8 safeguards do not detect any threats at all, despite they claiming to do so in their specifications.
Further, the majority of the safeguards do not identify common security threats.
Some of safeguards take between 45 seconds and 3 minutes to detect a security threat, and  
detection time tends to increase when the threat is more complicated. 
A key finding is that threats are not detected even when originating from the IoT device. This is concerning, given that recent works show that attacks launched by IoT devices are increasing~\cite{stiawan2019investigating,nawir2016internet,mohindru2020security,lyu2017quantifying}.

\subsection{Privacy Threats}

We now check if the safeguards detect any privacy threats.
Only 3 out of 8 safeguards claim to prevent/detect privacy threats: Avira, Bitdefender and Firewalla. 
Table~\ref{tab:evaluation} (Privacy) shows the results for the three privacy threats we test.

\noindent \emph{PII Exposure.} None of the safeguards detect PII exposure. 

\noindent \emph{Unencrypted Traffic.} This went undetected by all safeguards.

\noindent \emph{DNS Encrypted.} Only 2 out of 3 safeguards implement DoH when requested (Bitdefender and Firewalla). Avira does not implement DoH despite claiming to do so on its webpage.

\noindent \textbf{Takeaways}: Our results show that privacy threats are not detected by the safeguards under test, even if they claim to do so. Privacy threats are difficult to detect because PII vary by user, are not easily generalizable, and they require deep packet inspection (DPI)~\cite{fuchs2012implications}. The only privacy feature implemented is DoH. However, only 2 out of 3 safeguards do it properly.

\subsection{Time consistency}
We finally investigate whether our results are stable, i.e., they do not change over time. To verify this, we repeat our experiments at three different points in time over a period of three months. We compared the results from the three sets of experiments and verified that there are no differences. 
%, i.e., the protection offered by the safeguards did not increase or decrease over time.
 
\noindent \textbf{Takeaways}: 
The fact that the capability of detecting threats did not improve significantly over time might be a cause of concern when using safeguards for mitigating threats in dynamic IoT contexts.

%!TEX root = ../paper.tex
\section{Evaluation of Safeguard Side Effects}
\label{sec:rq3_evaluation}

In this section we answer our last research question by discussing what the side effects of the safeguards are. 
In this analysis, we use the methodology described in Section~\ref{sec:rq3} to investigate whether the protection policies implemented by the safeguards generate any side effects.
%In particular, we analyze, for each safeguard, the false positives, the overhead generated by its network traffic and the implication of its privacy policy.

\subsection{Overprotection}

In this analysis we check whether we get any ``false positive'' notifications over a period of one month.
We did not observe such notifications from any safeguards except for Bitdefender, which detected a possible scanning attack from a specific IP address towards one IoT device. 
%Figure~\ref{fig:overprotection} shows the notification of the prevented attack.
%The notification shows that the attack comes from a specific IP address.%\url{104.152.52.100}. 
%We further check the packets capture during that time by filtering by IP address. 
We discover that the traffic was coming from the Gateway and we manually confirmed it was not attack traffic, so we categorize the event as false positive and as example of overprotection. 

\noindent \textbf{Takeaways}: For the period of our study, most safeguards do not overprotect (i.e., they do not report threats that do not occur). Only Bitdefender reports a false positive threat, blocking a connection with a harmless destination, which may be useful for other purposes.

\subsection{Network Traffic Overhead}
\label{eval:traffic_over}

 \begin{figure}[t!]
  \centering
  \includegraphics[width=0.85\columnwidth]{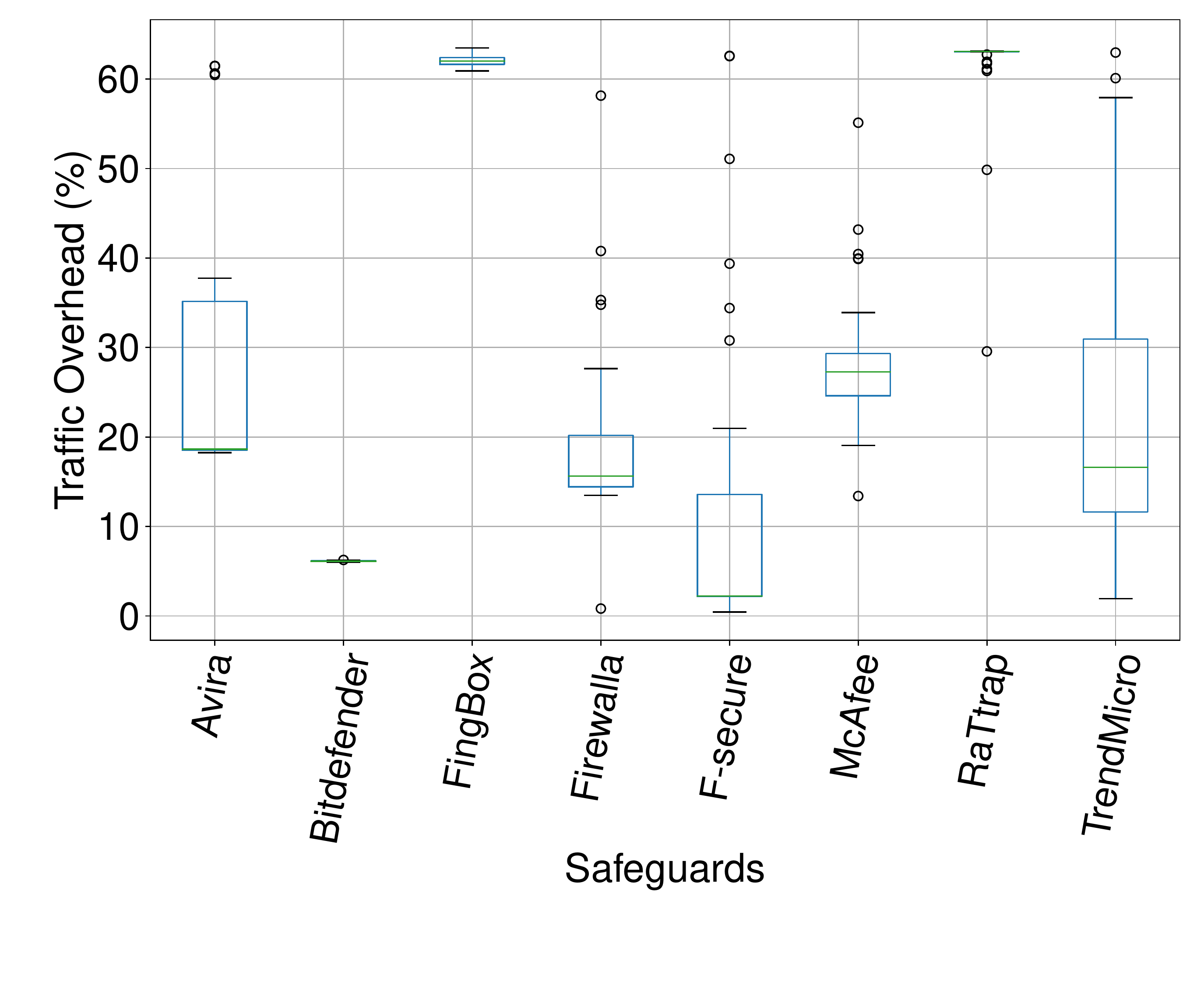}
    \vspace{-0.4cm}
  \caption{Average traffic overhead in percentage for each safeguard in 24 hours for a period of one month.}
  \label{fig:overhead}
\end{figure}

%In this section we compute the traffic overhead generated by the safeguards. 
We apply the methodology in Section~\ref{sec:rq3_traffic} to measure the percentage of the traffic produced by the safeguards over the total traffic.
Figure~\ref{fig:overhead} shows the average traffic overhead in percentage generated by the safeguards in 24 hours, over a period of one month.
Because the safeguards rely on the cloud, some users might be worried about traffic overhead.

Bitdefender and F-Secure have less traffic overhead than other safeguards. Avira, Firewalla, McAfee and TrendMicro have a traffic overhead that fluctuates between 20\% and 30\% of their traffic.
Fingbox and RATtrap have higher traffic overhead (more than 60\%).
Significant traffic is sent to the nearest M-Lab server, for speed test analysis.
Most of Fingbox traffic is due to ARP-spoofing (see Section~\ref{sec:safeguards}) and multicast/mDNS queries.
ARP-spoofing allows Fingbox to intercept network packets, but also
floods the home network---potentially causing slowdowns, faster battery depletion in wireless devices, and leads multiple devices to malfunction. 
Moreover, ARP-spoofing is recognized as malicious activity and some routers actively block it~\cite{kumar2010denial,kumar2005impact}.

\noindent \textbf{Takeaways}: Some of the safeguards introduce significant traffic overhead. In general the overhead is never less than 10\% of the traffic of the IoT devices.

\subsection{Privacy Policy} 
\label{sec:privacypolicy}

%!TEX root = ../paper.tex

\begin{table*}[t!]
        \centering
	\caption{Comparative analysis of the privacy policy for the safeguards under test}
	\vspace{-2mm}
	\label{tab:privacy_policy}
	\resizebox{1\textwidth}{!}{
		\centering
		\begin{tabular}{p{0.11\textwidth}||p{0.14\textwidth}|c|c|c|c|c|c|c}
		\centering
%			\textbf{Device} & \textbf{\#} & \textbf{\#} & \textbf{Non-required}\\ 
			\textbf{Privacy Policy} & \textbf{Avira~\cite{avirapp}} & \textbf{Bitdefender}~\cite{bitdefenderpp} & \textbf{F-Secure~\cite{fsecurepp}}& \textbf{Fingbox~\cite{fingboxpp}}& \textbf{Firewalla~\cite{firewallapp}}& \textbf{McAfee~\cite{mcafeepp}}& \textbf{RATtrap~\cite{rattrappp}}& \textbf{TrendMicro~\cite{trendmicropp}}\\ 
			\hline
			\hline
%General

Anonymization & \cmark& {\cmark} [pseudonymize]& {\xmark} [ceasing subscribtion]& \cmark& \xmark& \xmark& \xmark& \xmark   \\
\cline{1-9}
Usage of Personal Data & \cmark& \cmark& \cmark& \cmark& \cmark& \cmark& \cmark& \cmark   \\
\cline{1-9}
 Retention Period &In accordance with & 10 years& 6 months & As long  & Indefinitely& Subscription & Subscription & Ongoing legitimate   \\
&legal requirements & & & as necessary & & period&period & business need    \\
\cline{1-9}
Third Party &SaaS vendor, Akamai. Mixpanel, Ivanti &Partners &Partners & Partners& \xmark&Partners &Partners& Partners  \\
\hline
\hline
		\end{tabular}
	}
\end{table*}

In this section we study in detail the privacy policy for each safeguard and we align it to our findings. 
We focus on data anonymization, data usage, data retention period, and data sharing with third party entities. 
Table~\ref{tab:privacy_policy} shows the behavior of the safeguards with respect to those data policies. 
We report additional relevant text of the privacy policy for each safeguard in the Appendix.

\noindent \textbf{Anonymization.} 
%In this Section we analyze the privacy policy of the safeguards regarding the anonymization of personal data. 
%Table~\ref{tab:privacy_policy} shows that 
Only 3 safeguards out of 8 declare to anonymize personal data when retaining or sharing with others. However two of them contact directly third party entities so it is difficult to believe they can anonymize the IP address before sharing it with them. 
Avira states that they anonymize the data (See Table~\ref{tab:privacy_policy}), in particular they claim to ``transfer to their partners (Mixpanel, Akamai, Ivanti) only anonymized data.'' and ``The IP address is also anonymized as soon as possible.''  However, in our analysis in Section~\ref{sec:strategy}, we found that Avira is contacting Mixpanel directly (by contacting \url{api.mixpanel.com}). If the safeguard exchanges packets with the partner directly, the IP is not anonymized.
Bitdefender claims to ``apply adequate solutions to anonymize them, or at least to pseudonymize them.''
F-Secure anonymizes data only when the user ceases subscribing to their services. In that case ``the analytical data related to your service use will be reverted to anonymous data, and we are no longer able to associate it with you.''
Fingbox states that ``it is permitted to collect non-personal and anonymous data and anonymize personal data. This data does not allow Fing to proceed to identify an individual person and this data can be shared with third parties, including for statistical purposes.'' 
However, in our analysis in Section~\ref{sec:strategy}, we found that Fingbox is contacting directly the third party services Snapcraft and M-Lab, thus exposing the IP address with third party entities without anonymizing it.
Firewalla does not claim to anonymize personal data, but it informs the user to be aware of ``hackers that are constantly working to defeat security measures.''
McAfee, RATtrap and TrendMicro claim to use administrative, organizational, technical, and physical safeguards to protect the personal data they collect and process. Their security controls are designed to maintain data confidentiality, integrity, and an appropriate level of availability. However, there is no statement about anonymization of personal data.

\noindent \textbf{Usage of Personal Data.} 
%We now analyze the usage of personal data.
Most of the safeguards use personal data to make statistical analysis and market studies, operational processes, and data analysis.
They also store technical details for analysis, such as data for identifying the device (UUID), the infected URL the user reports or IP addresses.
F-Secure clearly states in its policy that suspicious or malicious sources and destinations are coming from an F-Secure-operated and -maintained system called \emph{Security Cloud}, confirming our finding in Section~\ref{sec:strategy}. %that F-Secure needs the cloud to operate.
F-Secure stores IP traffic source and destination addresses, DNS queries and replies per device MAC address(es), and traffic patterns. This possibly represents a privacy risk for the user.
Fingbox collects personal data (i.e., identification and contact data, connection data, including the IP address used to connect your device to the Internet, login information, browser type and version, time zone setting, browser plug-in types and versions, operating system and platform, type of device); however, they do not specify network traffic storage. 
Firewalla collects a range of personal information (See Table~\ref{tab:privacy_policy}). The Firewalla's IDS and IPS functions run inside the Firewalla box; however, portions of metadata will be sent to the cloud if needed, as demonstrated in Section~\ref{eval:traffic_over}. Firewalla inspects unencrypted portions of traffic, including the IP header, protocol headers, port numbers, domain name, duration of the flow, length transferred.
McAfee collects information for participating in threat intelligence networks, conducting research, and adapting products and services to help respond to new threats; they clearly indicate that they are using the cloud to process  threats.
RATtrap collects device identifiers, the domains/IP addresses classified as malicious, certain traffic metadata, and usage data. They indicate that they collect the IP address  to approximate the location. 
The stated purpose for data collection is research and development, and to market or promote products and services. 
TrendMicro collects personal information about the user, metadata of user/devices managed by gateway products, behaviors of products users, malicious connection information, network architecture and telemetry data.

\noindent \textbf{Personal Data Retention Period.} 
%Table~\ref{tab:privacy_policy} (third row) reports the data retention period declared by each safeguard. 
Avira, TrendMicro and Fingbox state they retain the data in accordance with legal requirements and ``as long as necessary.'' However, it is unclear what the legal requirements are. Other safeguards, such as Bitdefender and F-Secure indicate that they retain data for a limited period of time, while McAfee and RATtrap retain a user's account information for as long as the user's account is active.
More concerning, Firewalla stores some user information and content indefinitely.

\noindent \textbf{Third Party.}
%We now report data sharing practices with third-party entities.
Avira shares personal data with SaaS vendors and specific partners, such as Akamai. %, Mixpanel and Ivanti. 
Bitdefender, F-Secure, McAfee, Rattrap and TrendMicro share personal data with partners who perform services for them. McAfee shares aggregate data with third parties. Rattrap states that they may process and store user information outside of user country of residence, to wherever their or their third-party service providers operate for the purpose of providing the services.
TrendMicro states that they may also transfer personal information to their contractors based in various countries in the world
where they do business. They state that some of these countries may provide less legal protection than others for personal information.
Fingbox shares user data with selected third parties that can use the personal data only for the execution of their specific task (i.e., M-Lab for speed test). However, as stated by Fing, such recipients may be located in jurisdictions outside the European Economic Area that ``may not guarantee an appropriate level of personal data protection.''
Firewalla states that they do not sell or transfer to outside parties user personal information for commercial or marketing purposes.

\noindent \textbf{Takeaways}: Safeguard privacy policies present numerous concerns. Most user information is shared with third-party entities, sometimes without anonymization. These safeguards process/store network traffic and traffic patterns in the cloud, thus leaving the user vulnerable to privacy harms. Our analysis shows that sometimes the safeguards share data outside user's privacy jurisdiction and that even if some safeguards claim to share only anonymized IP addresses with third-party entities, they contact them directly, thus making the IP address visible anyway. 

%!TEX root = ../paper.tex
\section{Discussion}
\label{sec:discussion}

Our extensive experiments and results that cover a large number of safeguards, tested with many IoT devices, reveal a rather underwhelming outlook on the state of IoT security and privacy from commercial protection services today. In this section we briefly discuss some of the implications of our results, limitations of our investigations, and potential ways to move beyond the status quo. 

\noindent \textbf{Privacy and Security Implications.}
%Our results and findings have several privacy and security implications.
Our extensive evaluations show that safeguards might be providing their users with a  false sense of security and privacy, while not offering much protection in practice.
We observe that some safeguards establish connections and send data to servers outside the EU, a practice that is not compliant with the data localization requirements for the country where our testbed, devices, and associated user accounts are located. 
Another implication of the use of safeguards is that their privacy policies often indicate that they store data (in the cloud) only when malicious network traffic is detected; however, because they exhibit a non-zero number of false positives, they may mistakenly  store ``normal'' network connections from the user without informing them.
On a more alarming note, the privacy policies of some of these safeguards directly indicate personal data sharing with third parties. Home IoT data often include private and sensitive information and inferences~\cite{apthorpe2017smart}; hence, sharing these data can have undesired consequences for users.
Finally, some of the poor results we observed when analyzing existing safeguards could be due to conservative configurations that avoid false alarms and service outages; however, this design choice has the side effect of potentially exposing IoT users to undetected privacy and security risks. 

\noindent \textbf{Mitigation.}
Protecting against IoT security threats and privacy risks is a challenging task, due to the fragmented market, huge diversity of software and hardware providers, and lack of strong regulatory forces and market incentives. 
A shortcoming in many of today's safeguards is the use of rule-based methods, which are not scalable to the increasing diversity of devices and usages. 
The use of generic machine learning models and rule-based methods make the important problems of device identification and per-device behavior analysis rather challenging. 
\new{Possible mitigation could include the possibility to regularly train the ML models at the edge to keep up with the changes in device usage trends~\cite{kolcun2020case}.
To mitigate privacy issues, devices could focus on approaches that rely on local traffic analysis.}
Edge-based solutions running on the home gateway, such as IoTrim~\cite{mandalari-pets21}, have been recently proposed to mitigate shortcomings of cloud-based solutions. 
\new{We hope that our findings will encourage the industry to move in these directions.}

\noindent \textbf{Limitations.}
While we have made our best effort to investigate the important privacy and security measures offered by IoT safeguards, as a first attempt at this space, this work has a few limitations. 
One of them is that we have treated all safeguards equally and as a blackbox, without investigating or reverse engineering their code (often not available) or (undocumented) individual response strategies. 
Another limitation is that we can only claim that the safeguard did not detect the particular threats we tried, but we cannot generalize this to any other threats we did not try (even different instances of the same attacks with different parameters). While we can claim that the safeguards did/did not detect our particular threat instance, we cannot claim that they would not detect any other instance (due to multiple ways a single threat can be implemented). 
Lastly, the scalability of the threats is limited to the dozens of popular devices in our testbed, and we do not investigate all IoT devices on the market. 
Performing long-term attacks and longitudinal studies will be a valuable next step in future works building on our paper; hence we make all our code and data available with this paper at \url{https://iotrim.github.io/safeguards.html}. 
We also made our results available to the safeguards' manufacturers to encourage them to improve their approaches. 

\noindent \textbf{Feedback from vendors.}
Fingbox have responded to our findings. In general the response is consistent with what we found. Regarding the  IoT device identification, they clarify that Fingbox relies not only  on ARP, but on other protocols.  They clarify that \emph{ARP storm} is performed only on the specific device that the user wants to suspend or eliminate from the network. About open port detection, they clarify that vulnerability scanning is scheduled on a weekly basis or triggered manually by the user. In our study, the threat is not detected in either case. More details about the response are reported in the Appendix.
\new{We have confirmed with Bitdefender that their definition of anomalous behavior is consistent with ours.}
\new{We verified additional threats with the vendors. Bitdefender declares: ``PII Exposure would detect credentials sent in the clear, common default credentials, credit card data sent over HTTP, etc."}
\new{McAfee emphasizes that the analysis conducted on the D-Link router is powered by an old version and clarifies that a free subscription could have resulted in some features not being enabled on the D-Link router. However, at the time of the experiments, the free subscription was advertised as a complete solution.}

\noindent \textbf{Ethical Considerations.}
In our experiments we do not cause any real threat on the Internet. All experiments are contained within our own testbed. When conducting the experiments, we fully respected the ethical guidelines defined by our affiliated organization, and we received approval.

%!TEX root = ../paper.tex
\section{Related Work}
\label{sec:relatedwork}

A number of surveys and SoK papers have been recently published on the topic of IoT security and privacy for consumer IoT devices. We analyze in this section the contributions of these works with respect to our paper.

The increasing privacy and security risks in the consumer IoT market has led to several tools to protect against abnormal IoT traffic.  
SPIN~\cite{lastdrager2020protecting} is a tool to visualize and block IoT devices traffic. 
Mandalari et al.~\cite{mandalari-pets21} develop a system to identify and block non-essential IoT traffic. 
%It does so by triggering and probing IoT devices functions and determining \emph{non-required} destinations.
However, in this paper we focus on evaluating commercial IoT safeguards and systematically studying security and privacy threats. 

Alrawi et al. \cite{alrawi2019sok} carried out a security analysis
of smart home system components and reviewed common attack scenarios and mitigation techniques. While we recognize the importance of
%securing system components and highlight several relevant 
security mechanisms implemented at the device or
safeguards level, we evaluate commercial safeguards and perform active and automated experiments on them.

The survey from He et al.~\cite{he2021sok} discusses works on context sensing for access control in smart homes, and
pointed out flaws in existing systems under adversarial attacks. Context-sensing is out of scope for this paper.

Babun et al.~\cite{babun2021survey} perform an analysis of popular smart home platforms. The authors focus on commercial and open-source
platforms and compare their system and app programming models, communication protocols, third-party
components support, as well as point out their limitations when dealing with sensitive sensor data and apps.
In contrast, our study is about IoT safeguards. %offering a comprehensive study of them reacting to security and privacy threats.

Recently, Zavalyshyn at al.~\cite{zavalyshyn2022sok} examine 10 industrial and 37 academic smart home systems and compared their system
and threat models, as well as the ways they deal
with sensitive sensor and user data. However, they do not target IoT safeguards and their methodology does not allow them to systematically assess security and privacy threat reactions.

There are a number of existing tools for privacy risk analysis from IoT traffic. 
For examples, IoT Inspector~\cite{huang2019iot} collects smart home traffic in scale using ARP spoofing. 
In~\cite{moniotr}, the authors studied information exposure from 81 consumer IoT devices from two testbeds in different countries. 

\new{While IoT devices' security and privacy risks have been extensively studied, the effectiveness of the measures taken to reduce these risks has received little attention. 
The creation and publication of a methodology for assessing how well safeguards respond to common security threats and privacy risks is one of the paper's major contributions. This methodology can be used as a benchmark for additional research in this field and to pinpoint areas where current security measures need to be strengthened.}

%!TEX root = ../paper.tex
\section{Conclusion}
\label{sec:conclusion}

Protecting household IoT devices against security and privacy threats is an important ongoing challenge.
Commercial IoT security and privacy solutions and risk-mitigation safeguards are appearing in the market and being offered by ISPs and device providers, but there has been little-to-no insight into how---or how well---they work in practice.

In this paper, we took a quantitative approach in auditing these safeguards and their promises, as well as analyzed their data-collection and sharing practices for the first time.
We developed a scalable and automated methodology for evaluating the effectiveness of these safeguards against known IoT and network security attacks and threats.
We also investigated safeguards privacy protections, alongside manually inspecting their own privacy policies and practices.

Our extensive evaluations using a large variety of threats on several device categories on an advanced IoT testbed indicate underwhelming performance from commercially available safeguards. 
They often do not provide advertised protection; further, their data-sharing practices with their own clouds and other third parties might also introduce potential privacy threats to their users. 

Based on our findings, we argue there is need for continuous, independent auditing of security and privacy products in the IoT space to ensure they deliver on 
their promises. To assist with such efforts, we share our framework, datasets, and findings publicly to encourage better 
protection efforts and regulatory-compliant data sharing practices \new{at \url{https://iotrim.github.io/safeguards.html}.}

\section*{Acknowledgements}

We thank the anonymous reviewers and our shepherd Nick Nikiforakis for their constructive feedback. This work was supported by the EPSRC Open Plus Fellowship (EP/W005271/1), the EPSRC PETRAS (EP/S035362/1), the UKRI’s Strategic Priorities Fund under the SDTaP programme’s commercialization stream (10049005), and the NSF ProperData award (SaTC-1955227).

\balance

\bibliographystyle{plain}
%\interlinepenalty=10000
\bibliography{paper}

\appendix
%!TEX root = ../paper.tex
%\label{sec:appendix_policy}
\balance

In this appendix we report the response in detail from Fingbox, Bitdefender, McAfee, and for each safeguard, some 
relevant text from their privacy policy. Results about this study are reported in Section~\ref{sec:privacypolicy}.
%We focus on \emph{Anonymization}, \emph{Usage of Personal Data}, \emph{Personal Data Retention Period}, and \emph{Third party}.

\subsection{Feedback from Vendors}

\subsubsection{Fingbox Feedback}
%Fingbox has responded to our findings with some notes and clarifications. We report them in this Section.

\noindent \emph{About device recognition.}
Regarding ``Fingbox uses the content of the ARP packet to report the name of the IoT device. Bitdefender and F-secure use a more sophisticated methodology and report a more user-friendly name, such as Amazon Echo Speakers.'', we'd like to clarify that all Fing products rely on the Find Device Recognition, that uses a number of protocols data a --- such as ARP, Bonjour, UPnP, SNMP among them --- and machine learning to cluster and predict the model of the device. I'm not in the position to validate if only ARP was used to identify devices in your lab, but it's likely that more than one protocol was used.
The technology is used in all Fing products and by several other major manufacturers of network-aware devices
(\url{https://embedtech.lansweeper.com/products/device-recognition/}). The Systems and methods for determining characteristics of devices on a network has also been patented in US and EU (\url{https://patents.google.com/patent/US20190363943A1/en}). \\
\emph{About ARP-spoofing.}
In order to have a truly plug-and-play solution that doesn't require re-wiring the home setup, Fingbox adopts the technique of ARP spoofing to perform a few operations. It is to be noted and clarified that Fingbox doesn't perform the operation constantly --- which would make it burdensome for the network --- but instead applies it only on the specific targets intended for blocking access, pausing the internet and analyzing bandwidth's top users for troubleshooting. This can be verified through standard traffic recording.
In other words, we do not make the Fingbox a replacement of the router through an ``ARP storm'', but we act like it only on the specific targets that the user wants to suspend or eliminate from the network, with a conscious decision. Anti-ARP spoofing functionality for home commercial routers represent the exception as of today. \\
\emph{About Open port detection.}
Fingbox runs an internal (UPnP, NAT-PMP) inspection and external (Penetration test) scheduled, currently on a weekly basis, although the users have the possibility to run it also manually. Is the result of the test mentioned in the paper the outcome of the manual test, or the lack of a notification? \\
\emph{General.}
In general, Fingbox aims at informing user proactively about intrusion (new devices) in the network, informing them about what device they are and optionally blocking them automatically, forming a more effective journey of Access-Control-Lists widely available on routers but only used for slowly-evolving networks.
There is also no mention on Wi-Fi intrusion detections in the paper; we believe that the widespread usage of home WI-FI and the relative easiness of Wi-Fi hacking (de-authentication flood, evil twins, KRAK attacks) are a top priority for the target audience; that is why the Fingbox has dedicated antennas to monitor constantly the type of wireless activities.

\subsubsection{Bitdefender Feedback}
%\noindent Bitdefender also replied to our findings with some clarifications. 
\new{
\emph{General.}
The following detections should successfully complete in real world scenarios. We also noticed that extremely popular real world attacks, such as command injection and brute force, which Bitdefender BOX covers as well, do not appear to have been the scope of your team in this research. These could make a valuable follow-up, if not practical in the current paper. \\
\emph{Anomaly Detection} implies a period of up to 30 days of learning and may include previous knowledge from other devices. It's possible the test didn't allow sufficient data capture before evaluating. While we do understand the logic behind the ON/OFF test, we did not consider this a flag to trigger AD in the present implementation. \\
\emph{Weak Password} tests would probably be best if focused on Telnet, as most devices employ the protocol, not FTP. Granted, adding FTP scanning would not be superfluous. \\
\emph{PII Exposure} would detect credentials sent in the clear, common default credentials, credit card data sent over HTTP. \\
\emph{UDP Flooding} detection kicks in after around 10 minutes and covers ports 80 and 443. The method does not intercept UDP traffic, but rather interpret ICMP unreachable incoming packets for at least 10 minutes. (Our understanding of the methodology suggests your team waited some 20 minutes, but we assume the attack lasted less than 10 minutes, hence detection did not occur). \\
\emph{Port Scanning} works exceptionally well in the real world. It assumes ports are forwarded to the device. Our guess for no detection is the test pinged less than 5 ports using Nmap. \\
\emph{OS Scanning}, as implemented, would be implied using Port Scanning detection. If the test was done on the local network it's possible traffic never reached Bitdefender BOX, hence detection was not possible. \\
\emph{Unencrypted Traffic} blocking is impractical due to the sheer number of IoT devices that communicate this way. Granted, a UI flag informing the user would probably be desirable.
}

\subsubsection{McAfee Feedback}
\new{
The comparative study has been conducted on a D-Link router which is powered by a four-year-old McAfee Secure Home Platform solution version. In the last four years, McAfee Secure Home Platform solution has evolved significantly and enhanced its capability in the following areas: 

\emph{Device Identification.} 
Latest solution uses number of protocols such as DHCP, bonjour, UPnP, User Agents, Ping and ML models for device identification; Device de-duplication due to randomized MAC.  \\
\emph{Security Threat detection.}
Latest solution can detect the threats even when devices uses DoH/DoT channel for communication; Support for iCloud Private Relay; Malicious Detection using McAfee Threat Intelligence system.  \\
\emph{Privacy.} 
Latest solution supports DNS over HTTPS (DoH) and DNS over TLS (DoT); Anonymization support (Appendix) – Users have the option to make a request to customer support to anonymize their data in our system.  \\
Looking at the ``Table-1 Summary of the Characteristics of the Safeguards'', it appears that a free subscription is opted by the user. When the free subscription is opted, some/majority of the capabilities/features might not have been enabled on the D-Link router used for the analysis as it is purely depending on the subscription model which the user has opted for. For example, Malicious Detection, Ad Tracker, Vulnerability scan, Weak password, DDOS detection, IoT Anomaly detection etc. are not available in the free subscription.
In the real world, other router manufacturers who are using the latest McAfee Secure Home Platform solution (with all capabilities enabled) work exceptionally well in detecting threats mentioned in the analysis scope. However, at present port scanning and OS scanning is not in scope for D-Link partner.
 }

\subsection{Privacy Policy: Anonymization}

\noindent \emph{Avira.} Only anonymized data is transferred. The IP address is also anonymized as soon as possible. \\
\emph{Bitdefender.} For this purpose, we collect only that personal data absolutely necessary for the specified purposes, on a best efforts basis. We do not sell your data. For the collected information and data, we strive to apply adequate solutions to anonymize them, or at least to pseudonymize them.'' \\
\emph{F-Secure.} In our data analytics activities, we combine analytics data with the service data. The resulting combined data set then continues to be processed based on a ``legitimate interest.'' The previously collected analytical data is retained as part of the service statistics, as its retroactive removal would break the statistics. When you cease subscribing to our services, the analytical data related to your service will be reverted to anonymous data, we are no longer able to associate it with you. \\
\emph{Fingbox.} Fing is permitted to collect non-personal and anonymous data and anonymize personal data. This data does not allow Fing to proceed to identify an individual person and this data can be shared with third parties, including for statistical purposes. \\
\emph{Firewalla.} We use appropriate physical, electronic, and other procedures to safeguard and secure the information we collect. However, please be aware that the Internet is an inherently unsafe environment, and that hackers are constantly working to defeat security measures. Thus, we cannot guarantee that your information will not be accessed, disclosed, altered or destroyed, and you accept this risk. \\
\emph{McAfee.} Our security controls are designed to maintain confidentiality, integrity, and a level of availability. \\
\emph{RATtrap.} While we implement safeguards designed to protect your information, no security system is impenetrable and due to the inherent nature of the Internet, we cannot guarantee that data, during transmission through the Internet or while stored on our systems or otherwise in our care, is absolutely safe from intrusion by others. \\ 
\emph{TrendMicro.} Trend Micro has taken appropriate security measures --- including administrative,
technical (e.g., encryption, hashing, etc.) and physical measures --- to maintain and protect visitors'
personal information against loss, theft, misuse, unauthorized access, disclosure, alteration and
destruction. Access to visitors' personal information is restricted to authorized personnel only.

\subsection{Privacy Policy: Usage of Personal Data}

\noindent \emph{Avira.} Contract performance, Revenue generation, Operational processes, Marketing, Data analysis. \\
\emph{Bitdefender.}  To make statistical analysis and market studies; To perform marketing activities for Bitdefender's own needs.
Technical details, such as data for identifying the device (UUID), the infected URL you reported or an IP addresses. \\
\emph{F-Secure.} In addition to such specific purposes, the following general purposes of personal data use apply across all of our services:
Provisioning of services. To deliver our services to you, we process the data for the following purposes:
Customer journey. To identify authorized users, process and track transactions, administer user accounts, as well as for shipping, invoicing, and managing licenses.
Deliver, fix, and enhance. Delivering, maintaining, and developing our services and websites, and to provide help and support for the services.
Analyze, to track that our services are taken into use and how they are used so that we can improve the services, manage your customer relationship, and approach you with relevant messages.
Communicate. To send you information relating to the services, conduct customer surveys, market our services to you. The actual communication may be handled by F-Secure or by partners. \\
%Regulatory. To prevent fraudulent, illegal, or infringing activities and to comply with legal or regulatory requirements.
\emph{Fingbox.} We will collect your name, email address, location, password. This information will be combined with technical information about your device and other information we may collect about you or your use of the Services, and we will handle all information as described in this Policy. \\
\emph{Firewalla.} This information may include a device identifier, user settings, and the type of operating system used by your device. We may also collect information about your use of our Service.  location. We may do this by converting your IP address into a rough geo-location or access your mobile device's GPS coordinates if you enable location services on your mobile device. To provide you with Services, To improve our App and Services, For marketing purposes, To respond to your messages and comments, To send you technical notices, Firewalla's IDS and IPS functions run inside the Red or Blue Box , Portions of Meta Data will be sent to the cloud if needed.  See the Cloud section, Unless specified, all data remain local on the Firewalla Box, Firewalla Box only looks at the unencrypted portion of the traffic, IP Header, Protocol Headers (TCP, https, ssh, etc.), Port Numbers, Domain Name, Duration of the flow, Length transferred (upload / download). \\
\emph{McAfee.} Details about your computers, devices, applications, and networks, including internet protocol (IP) address, cookie identifiers, mobile carrier, Bluetooth device IDs, mobile device ID, mobile advertising identifiers, MAC address, IMEI, Advertiser IDs, and other device identifiers that are automatically assigned to your computer or device when you access the Internet, browser type and language, language preferences, battery level, on/off status, geo-location information, hardware type, operating system, Internet service provider, pages that you visit before and after using the Services, the date and time of your visit, the amount of time you spend on each page, information about the links you click and pages you view within the Services, and other actions taken through use of the Services such as preferences. We may collect this information through our Services or through other methods of web analysis. Details about your internet, app, or network usage (including URLs or domain names of websites you visit, information about the applications installed on your device, or traffic data); and performance information, crash logs, and other aggregate or statistical information. Analyze data sent to/from your device(s) to isolate and identify threats, vulnerabilities, viruses, suspicious activity, spam, and attacks, and communicate with you about potential threats;
Participate in threat intelligence networks, conduct research, and adapt products and services to help respond to new threats;
Encrypt your data, lockdown a device, or back-up or recover your data.
Check for Service updates and create performance reports on our Services, to ensure they are performing properly; and
Look for misuses of your data when you use our identity monitoring products. Authenticate your identity and prevent fraud with your biometric data;
Analyze your behavior to measure, customize, and improve our Site and Services, including developing new products and services;
Advertise McAfee products and services that we think may be of interest to you;
Establish and manage accounts and licenses, including by collecting and processing payments and completing transactions;
Provide customer support, troubleshoot issues, manage subscriptions, and respond to requests, questions, and comments;
Communicate about, and administer participation in, special events, programs, surveys, contests, sweepstakes, and other offers and promotions;
Conduct market and consumer research and trend analyses;
Enable posting on our blogs, forums, and other public communications;
Perform accounting, auditing, billing, reconciliation, and collection activities;
Prevent, detect, identify, investigate, and respond to potential or actual claims, liabilities, prohibited behavior, and criminal activity; and
Comply with and enforce legal rights, requirements, agreements, and policies. \\
\emph{Rattrap.} We collect information through your device about your device identifier, the domains you visit which we classify as malicious, the IP address you connect to which we classify as malicious, the IP addresses that attempt to connect to your router, and certain traffic metadata. We use your IP address in order to approximate your location to provide you with a better Service experience. How much of this information we collect depends on the type and settings of the device you use to access the Services.
Debugging Information.
We may request log files from your device or your mobile applications for the purpose of troubleshooting or debugging problems. These log files may contain traffic data, information about the operation of your device, information about your usage of mobile application and other data that can help us better diagnose your issues.
For research and development. To market or promote products.\\
\emph{TrendMicro.} Name, Phone number, Email address, Device ID, Operating system, License Key, Product information, such as MAC address, device ID, Public IP address of the user's gateway to the internet, Mobile/PC environment, Metadata from suspicious executable files, URLs, Domains and IP addresses of websites visited and DNS data, Metadata of user/device managed by gateway Product, Information about the Android applications installed on a user's device, Application behaviors, Personal information contained within email content or files to which Trend Micro is provided access, Behaviors of Product users, Information from suspicious email, including sender and receiver email address, and
attachments, Detected malicious file information including file name and file path, Detected malicious network connection information, Debug logs, Network Architecture/Topology and network telemetry data, Screen capture of errors, Windows event log content, WMI event content, Registry data.

\subsection{Privacy Policy: Personal Data Retention Period} 

\noindent \emph{Avira.} In accordance with legal requirements. \\
\emph{Bitdefender.} These data are being stored for a limited period, depending on its usefulness for the information security needs. Based on the current speed of technology, we will not need them for over 10 years from the day of the collection. \\
\emph{F-Secure.}  The information is stored as long as the respective support case remains unsolved. Once solved, the information is gradually deleted or anonymized within two years from closing the case.for the duration of an active service subscription plus for the grace period of six months thereafter. \\
\emph{Fingbox.} Your personal data will be retained as long as necessary for achieving the purpose for which they were collected and in line with the legal, regulatory and internal requirements in this respect. IP address used to connect your device to the Internet, your login information, browser type and version, time zone setting, browser plug-in types and versions, operating system and platform, type of device;
information on MAC addresses and Wi-Fi networks. \\
\emph{Firewalla.} We intend to store some of your information and User Content indefinitely. \\
\emph{McAfee.} McAfee will keep your Personal Data for the minimum period necessary for the purposes set out in this Notice, namely (i) for as long as you are a registered subscriber or user of our products or (ii) for as long as your Personal Data are necessary in connection with the lawful purposes set out in this Notice, for which we have a valid legal basis or (iii) for as long as is reasonably necessary for business purposes related to provision of the Services, such as internal reporting and reconciliation purposes, warranties or to provide you with feedback  you might request. \\
\emph{Rattrap.} We retain your account information for only as long as your account is active.
Connection Information.
We retain information collected from your devices about blocked domains and blocked IP addresses for a period of thirty (30) days. We retain information derived from cookies and other tracking technologies for a reasonable period of time from the date such information was created.After such time, we will either delete or anonymize your information. \\
\emph{TrendMicro.} Trend Micro will keep visitors' personal information for as long as we have an ongoing legitimate business 
need to do so (for example, to provide a service a visitor has requested or to comply with
applicable legal, tax or accounting requirements).
When we have no ongoing legitimate business need
to process visitors' personal information, we will either delete or pseudonymize it or, if this is not possible
(because visitors' personal information has been stored in backup archives), then we will
securely store that personal information until deletion is possible.

\subsection{Privacy Policy: Third Party}

\noindent \emph{Avira.} Active group employees, SaaS vendor. Legitimate forwarding of personal data to Akamai, Mixpanel, Ivanti. \\
\emph{Bitdefender.} Bitdefender may allow limited access to its Partners. Access will be allowed only to certain data related to its referred clients and just for fulfilling the contractual obligations between Bitdefender and its Partner for selling or for support of Bitdefender products. \\
\emph{F-Secure.} We exchange (both disclose and receive) some of your personal data with our distribution partners (operators, webstores, etc.), who market, distribute, administer, and support our services. We may transfer or disclose some of your personal data to F-Secure group companies and our subcontractors who help us create the services.We also exchange with the partner such above listed data (e.g. status of your subscription, installation success, service in active use, data collected for resolving a technical support case) as is necessary and proportional.We have to exchange some data (such as ``Android marketing identifier'' and other like identifiers) with our online analytics and marketing partners to enable our digital analytics and marketing activities. The vast majority of the data is not shared with others. \\
\emph{Fingbox.} In principle we do not share your personal data with anybody outside Fing. Sometimes it is however necessary that for operational reasons we appeal to meticulously selected third parties within the scope of the purposes described in article 4. These recipients can only use the indicated personal data for the execution of their specific task, need to act in respect with our privacy statement and they cannot use them for any other purposes. Such recipients may be located in jurisdictions outside the EE Area that may not guarantee an appropriate level of data protection. \\
\emph{Firewalla.} We do not sell, trade, or otherwise, transfer to outside parties your Personally Identifiable Information (PII) for commercial or marketing purposes. \\
\emph{McAfee.} With current and future members of the McAfee family of companies for the purposes described in this Notice;
With service providers who perform services for us (see the list of our sub-processors for Consumer Products is available upon request for where required under applicable laws).
If we believe disclosure is necessary and appropriate to prevent physical, financial, or other harm, injury, or loss, including to protect against fraud or credit risk;
To legal, governmental, or judicial authorities as instructed or required by those authorities and applicable laws, or in relation to a legal activity, such as in response to a subpoena or investigation of suspected illicit or illegal activities, or where we believe in good faith that users may be engaged in illicit or illegal activities, or where we are bound by contract or law to enable a customer or business partner to comply with applicable laws;
In connection with, or during negotiations for, an acquisition, merger, asset sale, or other similar business transfer that involves all or substantially all of our assets or functions where Personal Data is transferred or shared as part of the business assets (provided that such party agrees to use or disclose such Personal Data consistent with this Notice or gains your consent for other uses or disclosures);
With your consent or at your direction, such as when you choose to share information or publicly post content and reviews (for example, social media posts); and
With persons of your choosing and at your discretion, should the product you are subscribed to allow that functionality.
We may also share aggregate data that does not identify you or any specific device with third parties. \\
\emph{RATtrap.} We collect information globally and primarily store that information in the United States. We transfer, process and store your information outside of your country of residence, to wherever we or our third-party providers operate for the purpose of providing you the Services. \\
\emph{TrendMicro.} Trend Micro may share personal information with its affiliated companies,
resellers, distributors, vendors, service providers or partners in order to provide the high quality,
localized Products or offers that Customers have requested, and/or to meet Customer needs or provide
support. Trend Micro may engage contractors to provide certain services in connection with Products,
such as providing technical support, hosting cloud services, handling order processing or shipping
Products, conduct Customer research or satisfaction surveys.Trend Micro may also disclose personal information if we determine that disclosure is necessary to
enforce our terms and conditions or protect our Products.

\end{document}